\documentclass[12pt]{iopart}

\usepackage{iopams}  
\usepackage{graphicx}
\usepackage{amssymb}
\usepackage{bm, bbm}    
\usepackage{bm}    
\usepackage{color}
\usepackage{subfigure}
\usepackage{multirow} 
\usepackage{array} 

\begin{document}

\title{Protected boundary states in gapless topological phases}

\author{Shunji Matsuura$^1$,
  Po-Yao Chang$^2$,
  Andreas P. Schnyder$^3$,
  Shinsei Ryu$^2$}

\address{$^1$Departments of Physics and Mathematics, McGill University, Montr\'eal, Qu\'ebec, Canada}
\address{$^2$Department of Physics, University of Illinois at Urbana-Champaign, 1110 West Green St, Urbana, IL 61801, USA}
\address{$^3$Max-Planck-Institut f\"ur Festk\"orperforschung, Heisenbergstrasse 1, D-70569 Stuttgart, Germany} 

\ead{
 \mailto{a.schnyder@fkf.mpg.de}
\mailto{chang153@illinois.edu}}

\date{\today}

\begin{abstract}
We systematically study gapless topological phases of \mbox{(semi-)}metals and nodal superconductors described
by Bloch and Bogoliubov-de Gennes Hamiltonians.
Using K-theory, a classification of topologically stable Fermi surfaces in 
\mbox{(semi-)}metals and nodal lines in superconductors is derived.
We discuss a generalized bulk-boundary correspondence that relates the topological features of the Fermi surfaces and superconducting nodal lines to the presence of protected zero-energy states at the boundary of the system. Depending on the case,
the boundary states are either linearly dispersing (i.e., Dirac or Majorana states)
or are dispersionless, forming  two-dimensional surface flat bands or one-dimensional arc surface states.
We study examples of gapless topological phases
in symmetry class AIII and DIII, focusing in particular
on nodal superconductors, such as nodal noncentrosymmetric superconductors.
For some cases we explicitly compute the surface spectrum and examine the signatures of the topological boundary states in the surface density of states. We also discuss the robustness of the surface states against disorder.

\end{abstract}



\submitto{\NJP}
\maketitle

\section{Introduction}

The recent discovery of
topological electronic phases in insulating materials with strong spin-orbit 
coupling
\cite{konig07,
  hsiehNature08,
  hasanKaneReview10,
  qiZhangReview10,
  HasanMooreReview11,
  ryuNJP10} 
has given new impetus to the investigation of topological phases of matter.
Topological materials, such as 
the integer quantum Hall state and the spin-orbit induced topological insulators, are characterized by a nontrivial band topology, which gives rise to  protected exotic edge (or surface) states.
Many interesting phenomena,
including magneto-electric effects~\cite{zhangPRB08} and the emergence of localized Majorana states~\cite{fuKanePRL08},
have been predicted to occur in these systems. These phenomena could potentially lead to a variety of new technical applications, including novel devices for spintronics and quantum computation.

Besides the topological insulators and the integer quantum Hall state, which have a full bulk gap, there are also gapless phases that belong to the broad class of topological materials,
such as, e.g., (semi-)metals with topologically protected Fermi points and nodal superconductors with topologically stable nodal lines.
These gapless topological phases also exhibit
exotic zero-energy edge (or surface) states with many interesting properties. These boundary states
may be linearly dispersing (i.e., of Dirac or Majorana type),
or dispersionless, in which case they form
either two-dimensional surface flat bands or one-dimensional arc surface states. 
Notable examples of gapless topological materials include,
among others~\cite{heikkila2011a,heikkila2011b,volovik2011c},
graphene~\cite{nakada96,fujita96,neto2009,Niimi2005,Kobayashi2005,Niimi2006},
$d_{x^2-y^2}$-wave superconductors~\cite{huPRL94,ryu2002,kashiwaya2000,Kashiwaya1995,Alff1997,Wei1998},
the A phase of superfluid ${}^3$He~\cite{volovikJETP2011,tsutsumi2011}, 
and nodal
noncentrosymmetric superconductors~\cite{WangFa2012,sato06,beriPRB2010,Schnyder2010,brydon2010,Schnyder2011,satoPRB2011,yadaPRB2011,tanakaPRL2010,satoPRL2010}.

The topologically stable Fermi points and superconducting nodal structures in the aforementioned materials can be viewed, in a sense, as momentum-space defects, that is, as momentum-space analogues of real-space topological defects.
In other words, the nodal points in $d_{x^2-y^2}$-wave superconductors, the Fermi points in graphene, and the nodal points in ${}^3$He A can be interpreted as  momentum-space point defects, i.e., as vortices and hedgehogs, respectively. The nodal lines in noncentrosymmetric superconductors, on the other hand, correspond to momentum-space line defects, i.e., vortex lines.
Similar to real-space defects, the  stability of these Fermi points, nodal points, and nodal lines is guaranteed by the conservation of some topological invariant, i.e., e.g., a Chern or winding number.

In this paper, building on previous works~\cite{volovikJETP2011,sato06,beriPRB2010,Horava,volovik2007,BernardKimLeClair1,BernardKimLeClair2,
  schnyderPRB08,zirnbauerMathPhys96,altlandZirnbauer97,heinznerCommMath05,kitaev09,Teo-Kane2010,WanTurnerAschvin}, 
we derive a classification of topologically stable Fermi surfaces in \mbox{(semi-)}metals and nodal lines in superconductors using K-theory arguments (Table~\ref{single fermi} in Sec.~\ref{sec:LocStab})
\footnote{
By definition, a Fermi surface is a set of gapless points in the Brillouin zone.
To simplify terminology, we will refer to Fermi points/lines in metals and nodal points/lines in superconductors, 
etc., simply as  ``Fermi surfaces".}
\footnote{
Ho\v{r}ava~\cite{Horava} pointed out an interesting connection between
the classification of stable Fermi surfaces
and the classification of stable D-branes.
Making an analogy with string theory,
we observe that the result by Ho\v{r}ava
 corresponds to D$p$-branes in Type IIA string theory. 
Hence, one might wonder what 
are the gapless topological objects in condensed matter physics that
correspond to D$p$-branes in Type IIB string theory.
Furthermore, in Type I or Type I' string theory it is customary to consider besides
D$p$-branes also orientifold hyperplanes.
(Note that every space-time point on an 
orientifold hyperplane
is identified with its mirror image.)
It is known that D$p$-branes in Type I string theory are classified in terms of
real K-theory \cite{Witten:1998cd}.
Hence, one might again
wonder what are the gapless topological objects in condensed matter physics that 
correspond to orientifold hyperplanes or D$p$-branes 
in Type I or Type I' string theory.
In fact, for topological insulators and superconductors,
it was found that there is a one-to-one correspondence 
between the K-theory classification of topological insulators/superconductors 
and the K-theory charges of D-branes
in Type IIA and Type IIB string theory, or the K-theory charges of
non-BPS D-branes
in Type I and I' string theory~\cite{DbranesTI, DbranesTIPRD}.
}.
As it turns out,
the presence of discrete symmetries, such as time-reversal symmetry (TRS) or particle-hole symmetry (PHS), plays a crucial role
in the classification of gapless topological phases, a fact that has not been emphasized previously.
The appearance of protected zero-energy states at the boundary of gapless topological phases is discussed, and it is shown
that the existence of these boundary states is directly linked to the topological stability of the Fermi surfaces (superconducting nodal lines) in the bulk via a generalized bulk-boundary correspondence (Sec.~\ref{bulkBndCorr}). In particular, we demonstrate that 
gapless topological phases in symmetry class A or AIII with stable Fermi surfaces of codimension $p=d_k+1>1$
necessarily support zero-energy surface flat bands.
Finally, in Sec.~\ref{ss classes of gapless modes}, we present a few examples of gapless topological phases and discuss their topological surface states.

\section{Local stability of Fermi surfaces}
\label{sec:LocStab}

The classification of topologically stable Fermi surfaces in terms of K-theory is closely related
to the classification of topologically stable zero modes localized on real-space defects. 
In Sec.~\ref{real-defect TeoKane}, we will therefore first review the stability of localized gapless modes on topological defects, 
before discussing the classification of topologically stable Fermi surfaces in Sec.~\ref{topFSs}.
To denote the dimensionality of the Brillouin zone (BZ), the Fermi surfaces, and the real space defects we use
 the notation:
\begin{eqnarray}
d_{\mathrm{BZ}}
&=&
(\mbox{total spatial dimension})
\nonumber \\
&=&
(\mbox{total dimension of the BZ}), 
\nonumber \\
d_k
&=&
(\mbox{codimension of a Fermi surface})-1
\nonumber \\
&=&
\left(
\begin{array}{c}
\mbox{$\#$ of parameters characterizing a surface} \\
\mbox{surrounding a Fermi surface in the BZ}
\end{array}
\right), 
\nonumber \\
d_r
&=&
(\mbox{codimension of a real space defect})-1
\nonumber \\
&=&
\left(
\begin{array}{c}
\mbox{$\#$ of parameters characterizing a surface} \\
\mbox{surrounding a real space defect}
\end{array}
\right).
\nonumber
\end{eqnarray}
In other words, the dimension of a Fermi surface and a real space defect are $d_{\mathrm{BZ}}-d_k-1$ and $d_{\mathrm{BZ}}-d_r-1$, respectively.

\subsection{Real-space defects}
\label{real-defect TeoKane}

In this subsection, 
we review the classification of localized gapless modes on topological 
defects from the K-theory point of view \cite{Teo-Kane2010,Freedman2010}.
To that end, let us  consider the topology associated with gapped
Hamiltonians $\mathcal{H}(r,k)$, where 
$k=(k_1,k_2,\cdots, k_{d_{\mathrm{BZ}}})$ denotes the $d_{\mathrm{BZ}}$-dimensional momentum coordinate,
and $r=(r_1, r_2, \cdots, r_{d_r})$  the position-space parameters characterizing the defect.
That is, $r$ are the coordinates parametrizing
the surface
that encloses the defect in question.
For instance, 
 a line defect in a three-dimensional system is described by the Hamiltonian
$
\mathcal{H}(r,k)
=k_1\gamma_1+k_2\gamma_2+k_3\gamma_3
+m_1(x,y)\gamma_4+m_2(x,y)\gamma_5 
$,
where
$m_1(x,y)=x/\sqrt{x^2+y^2}$,
$m_2(x,y)=y/\sqrt{x^2+y^2}$, 
and $\gamma_i$ are five $4\times 4$ anticommuting matrices.
In this case, $k_{1,2,3}\in k$ and $m_{1,2} \in r$.
\begin{table}[t!]
\begin{center}
\begin{tabular}{cc | cccccccccccccc} 
\multicolumn{11}{l}{\textbf{complex case ($\mathbb{F}=\mathbb{C}$):}} \\
\hline
\multicolumn{2}{c|}{Symmetries} &  \multicolumn{9}{c}{$\delta = d_2 -d_1$} \\  
$s$& class    & $0$ & $1$ & $2$ & 
$3$ & $4$ & $5$ & $6$ & $7$ & $\cdots$ \\ \hline\hline
0 &A   & $\mathbb{Z}$ & 0
    & $\mathbb{Z}$ & 0 & $\mathbb{Z}$ & 0
    & $\mathbb{Z}$  & 0 & $\cdots$
\\\hline
1&AIII  & 0 & $\mathbb{Z}$ & 0 & $\mathbb{Z}$
      & 0 & $\mathbb{Z}$ & 0 & $\mathbb{Z}$
& $\cdots$
\\\hline
\end{tabular}
\end{center}
\vspace{0.5cm}
\begin{center}
\begin{tabular}{cc | cccccccccccccc}
\multicolumn{11}{l}{\textbf{real case ($\mathbb{F}=\mathbb{R}$):}} \\ 
\hline
\multicolumn{2}{c|}{Symmetries} &  \multicolumn{9}{c}{$\delta = d_2 -d_1$} \\  
$s$ & class & $0$ & $1$ & $2$ 
& $3$ & $4$ & $5$ & $6$ & $7$ 
& $\cdots$
\\ \hline\hline
0&AI  
    & $\mathbb{Z}$ 
    & 0 & 0 & 0 
    & $\mathbb{Z}$  & 0 & $\mathbb{Z}_2$ 
    & $\mathbb{Z}_2$ 
    & $\cdots$
    \\ \hline

1&BDI 
    & $\mathbb{Z}_2$ & $\mathbb{Z}$ 
    & 0 & 0 & 0 
    & $\mathbb{Z}$  & 0 
    & $\mathbb{Z}_2$ 
    & $\cdots$
    \\ \hline

2&D   

    & $\mathbb{Z}_2$ 
    & $\mathbb{Z}_2$ & $\mathbb{Z}$ & 0 
    & 0 & 0 & $\mathbb{Z}$ 
    & 0 & $\cdots$
    \\ \hline

3&DIII

    & 0 & $\mathbb{Z}_2$ 
    & $\mathbb{Z}_2$ & $\mathbb{Z}$ & 0 
    & 0 & 0 
    & $\mathbb{Z}$  & $\cdots$
    \\ \hline

4&AII

    & $\mathbb{Z}$  
    & 0 & $\mathbb{Z}_2$  & $\mathbb{Z}_2$ 
    & $\mathbb{Z}$ & 0 & 0 
    & 0& $\cdots$
    \\ \hline

5&CII 

    & 0 & $\mathbb{Z}$  
    & 0 & $\mathbb{Z}_2$  & $\mathbb{Z}_2$ 
    & $\mathbb{Z}$ & 0 
    & 0& $\cdots$
    \\ \hline
6&C
    & 0 
    & 0 & $\mathbb{Z}$ & 0 
    & $\mathbb{Z}_2$ & $\mathbb{Z}_2$ & $\mathbb{Z}$ 
    & 0 
    & $\cdots$
    \\ \hline
7&CI  
    & 0 & 0 
    & 0 & $\mathbb{Z}$ & 0 
    & $\mathbb{Z}_2$ & $\mathbb{Z}_2$  
    & $\mathbb{Z}$ 
    & $\cdots$
    \\ \hline
\end{tabular}
\end{center}
\caption{
K-group 
$K_{\mathbb{F}}(s;d_1, d_2)=K_{\mathbb{F}}(s;\delta=d_2-d_1)$
from
Ref.\ \cite{Teo-Kane2010}.
}
\label{periodic table of Fermi surf}
\end{table}

For condensed matter systems
defined on a lattice, the BZ is 
a $d_2$-dimensional torus,
$k\in T^{d_{\mathrm{BZ}}=d_2}$,
and $r\in S^{d_r=d_1}$, where $S^{d_1}$ is 
a $d_1$-dimensional sphere
surrounding the defect in real space. 
If we are interested in ``strong'' 
but not in ``weak''
topological insulators and superconductors,
we can take
$k\in S^{d_2}$. 
Furthermore, it turns out it is enough to consider 
$(r,k)\in S^{d_1+d_2}$ \cite{Teo-Kane2010}. 
To determine the topology of the family of Hamiltonians $\mathcal{H}(r,k)$,
one needs to examine the adiabatic evolution of the wavefunctions of $\mathcal{H}(r,k)$
along a closed real-space path surrounding the defect \footnote{It is assumed that the path is 
sufficiently far away form the singularity of the defect.}. From this consideration, one 
can define a K-theory charge for $\mathcal{H}(r,k)$
and describe the stable equivalent classes of Hamiltonians
$\mathcal{H}(r,k)$ in terms of the K-group 
\begin{eqnarray}
K_{\mathbb{F}}(s;d_1, d_2) ,
\end{eqnarray}
where 
$s$ represents one of the Altland-Zirnbauer symmetry classes~\cite{zirnbauerMathPhys96,altlandZirnbauer97,heinznerCommMath05} given 
in Table \ref{periodic table of Fermi surf},
 $\mathbb{F}=\mathbb{C}$ ($\mathbb{R}$) stands
for the complex (real) 
Altland-Zirnbauer symmetry classes, and
$d_1$ and $d_2$ represent
the dimensions of $r$
and $k$, respectively. 
An important relation used in the analysis of Ref.\ \cite{Teo-Kane2010,Freedman2010} is that 
K-groups of different symmetry classes are related by
\begin{eqnarray}
K_{\mathbb{F}}(s;d_1, d_2+1)
=
K_{\mathbb{F}}(s-1;d_1,d_2),
\label{tk result0}
\end{eqnarray}
and
\begin{eqnarray}
K_{\mathbb{F}}(s;d_1+1,d_2)
=
K_{\mathbb{F}}(s+1;d_1,d_2). 
\label{tk result1}
\end{eqnarray} 
Relations (\ref{tk result0}) and (\ref{tk result1}) can be derived by
considering smooth interpolations/deformations between two Hamiltonians  belonging to different symmetry classes and 
different position-momentum dimensions ($d_1$, $d_2$), thereby demonstrating that the two Hamiltonians
are topologically equivalent \cite{Teo-Kane2010,Freedman2010}.
Combining Eqs.~(\ref{tk result0}) and (\ref{tk result1}), one finds  
\begin{eqnarray}
K_{\mathbb{F}}(s;d_1+1, d_2+1)
=
K_{\mathbb{F}}(s;d_1,d_2),
\label{tk result2}
\end{eqnarray}
which shows 
that the topological classifications only depend on the difference
\begin{eqnarray}
\delta=d_2-d_1.
\end{eqnarray}
From this, it was shown in Refs.~\cite{Teo-Kane2010,Freedman2010} that the classification of zero-energy modes localized on real-space topological defects
is given by the K-groups $K_{\mathbb{F}}(s;d_1,d_2)$ (see Table \ref{periodic table of Fermi surf}) with
\begin{eqnarray}
d_1=d_r,
\quad
d_2=d_{\mathrm{BZ}}, 
\quad
\delta =d_{\mathrm{BZ}}-d_r. 
\end{eqnarray}
In other words, whenever the
 K-group is nontrivial 
 (i.e., $K_{\mathbb{F}}(s;d_r,d_{\mathrm{BZ}}) = \mathbb{Z}$ or $\mathbb{Z}_2$)
 the K-theory charge can take on nontrivial values, 
which in turn indicates the presence of one or several zero-energy modes
localized on the topological defect. 
As a special case, the periodic table of topological insulators
and superconductors~\cite{schnyderPRB08,kitaev09,2009AIPC.1134...10S,ryuNJP10} is obtained from the K-groups
 by taking (cf.\ Table~\ref{SumClass})
\begin{eqnarray}
d_1=0,
\quad
d_2=d_{\mathrm{BZ}}, 
\quad
\delta =d_{\mathrm{BZ}}. 
\end{eqnarray}

Representative Hamiltonians of the stable equivalent classes of  $\mathcal{H} ( r, k)$ can be constructed in terms of
linear combinations of anticommuting  Dirac matrices \cite{Teo-Kane2010} (see also~\cite{ryuNJP10}).
For instance, consider
\begin{eqnarray}
\mathcal{H}(r,k)=R^{\mu}(r,k) \gamma_{\mu}+K^{i}(r,k) \gamma_{i},
\end{eqnarray}
with ``position-type" 
Dirac matrices $\gamma_{\mu}$ and ``momentum-type" Dirac matrices $\gamma_i$,
where
$\{\gamma_{\mu},\gamma_{\nu}\}=2\delta_{\mu\nu}$, $\{\gamma_{i},\gamma_{j} \}=2\delta_{ij}$, and $\{\gamma_{\mu},\gamma_{i}\}=0$. 
If the Hamiltonian satisfies 
time-reversal symmetry $T$, we require
\begin{eqnarray}
[\gamma_{\mu},T]=\{\gamma_{i},T\}=0,
\end{eqnarray}
while for particle-hole symmetry $P$, we have
\begin{eqnarray}
\{\gamma_{\mu},P\}=[\gamma_{i},P]=0.
\end{eqnarray}
Under the antiunitary symmetries $T$ and $P$ the coefficients $R^{\mu} (r, k)$ and $K^i ( r , k)$  transform in the same way as position and momentum, respectively, i.e., 
$R^{\mu} (r, - k) = R^{\mu} (r,  k)$ and $K^i ( r , - k) = - K^i ( r , k)$.
As shown in Ref.~\cite{Teo-Kane2010}, a  representative Hamiltonian of the real symmetry class $s$ can be constructed in terms
of  a linear combination of $b+1$ position-type matrices $\gamma_{\mu}$ and 
$a$ momentum-type matrices $\gamma_{i}$, with 
$
  a-b=s
  \; \mbox{mod}
  \; 8
$.

\begin{table}
\begin{center}
\begin{tabular}{lccccccccccccc}\hline
& $d_1$ & $d_2$ & $\delta$ &
\\ \hline\hline
 $\begin{array}{l}
 \mbox{insulators (fully gapped SCs)}
 \end{array}$  &  0 & $d_{\mathrm{BZ}}$   & $d_{\mathrm{BZ}}$ 
    \\ \hline
$\begin{array}{l}
\mbox{zero modes localized on real-space} \\
\mbox{defects in  insulators (fully gapped SCs)}
\end{array}$
 & $d_r$ & $d_{\mathrm{BZ}}$  & $d_{\mathrm{BZ}} -d_r$
    \\ \hline
$\begin{array}{l}    
\mbox{Fermi surfaces}
\end{array}$   &  0 & $d_k$   & $d_k$   \\
\hline 
\end{tabular}
\end{center}
\caption{
Assignment of dimensions  ($d_1$, $d_2$) for different classification schemes:
(i) classification of  insulators (fully gapped superconductors),
(ii) classification of zero modes localized on real-space defects in insulators (fully gapped superconductors), and
(iii) classification of Fermi surfaces.  
\label{SumClass}
}
\end{table}

\subsection{Fermi surfaces (momentum-space defects)}
\label{topFSs}

The analysis of Refs.~\cite{Teo-Kane2010,Freedman2010},
which we have reviewed above, 
can be extended to study the topological stability of Fermi surfaces. 
For a given Hamiltonian $\mathcal{H} (k)$, we define the Fermi surface as the momentum-space manifold where $\mathcal{H} ( k ) =0$~\footnote{Alternatively, the Fermi surface can be defined in terms of the poles of the single particle Green's function.}.
The key observation is that topologically stable Fermi surfaces can be viewed
as defects in the momentum-space structure of the wavefunctions of $\mathcal{H} ( k)$.
Hence, in order to determine the topology of
a $(q=d_{\mathrm{BZ}}-d_k-1)$-dimensional Fermi surface,
we need to examine the adiabatic evolution of the wavefunctions of $\mathcal{H} (k)$  
along a closed momentum-space path surrounding the Fermi surface.
This closed path in momentum space is parametrized by $d_k$ variables, i.e., it defines
a $d_k$-dimensional hypersphere 
$S^{d_k}=S^{d_{\mathrm{BZ}}-q-1}\in S^{d_{\mathrm{BZ}}}=\mathrm{BZ}$ 
surrounding the Fermi surface. 
Hence, the topological stability of a $q$-dimensional Fermi surface in a $d_{\mathrm{BZ}}$-dimensional BZ
is describe by the K-group $K_{\mathbb{F}}(s;d_1,d_2)$, with 
\begin{eqnarray}
d_1 = 0,
\quad
d_2 = d_k,
\quad
\delta=d_k,
\end{eqnarray}
i.e., by
$
K_{\mathbb{F}}(s;\delta=d_k)
$, where $d_k = d_{\mathrm{BZ}} -q -1$.
That is, 
the  classification (or ``periodic table'')
of topologically stable Fermi surfaces in symmetry class $s$
can be inferred from Table \ref{periodic table of Fermi surf} 
together with Table~\ref{SumClass}.
For two- and three-dimensional systems, the classification
of $q$-dimensional Fermi surfaces is explicitly given in 
Table~\ref{single fermi}~\footnote{%
Focusing on noninteracting systems, we study
the topological stability of Fermi surfaces in terms of Bloch or Bogoliubov-de Gennes Hamiltonians.
However, it is straightforward to extend our analysis to the Green's function formalism describing weakly
(or moderately weakly) interacting systems (see Refs.~\cite{Grinevich88,Horava,Teo-Kane2010} and compare with Ref.~\cite{zhangPRX12}).}.

\begin{table}[t!]
\begin{center}
\begin{tabular}{c | ccc | cccccccccc}
\multicolumn{6}{l}{\textbf{complex case  ($d_{\mathrm{BZ}}=2$):}} 
\\ \hline 
\multirow{2}{*}{class}  & \multirow{2}{*}{$T$} & \multirow{2}{*}{$P$} & \multirow{2}{*}{$S$}  & $d_k=0$ & $d_k=1$ 
\\
& &  & & line & point 
 \\ \hline\hline
A & $0$ & $0$ & $0$ & $\mathbb{Z}$ & 0
\\ \hline
AIII & $0$ & $0$ & $1$ & 0 & $\mathbb{Z}$ 
\\\hline
\end{tabular}
\hspace{0.5cm}
\begin{tabular}{c | ccccccccccccc}
\multicolumn{4}{l}{\textbf{complex case ($d_{\mathrm{BZ}}=3$):}  }\\ \hline 
\multirow{2}{*}{class}    &    $d_k=0$& $d_k=1$ & $d_k=2$
\\ 
  & surface & line & point 
     \\ \hline\hline
A   & $\mathbb{Z}$ & 0
    & $\mathbb{Z}$ 
\\\hline
AIII  & 0 & $\mathbb{Z}$ & 0 
\\\hline
\end{tabular}
\end{center}
\vspace{0.5cm}
\begin{center}
\begin{tabular}{c | ccc | cccccccccc}
\multicolumn{6}{l}{\textbf{real case  ($d_{\mathrm{BZ}}=2$):}} \\ \hline
\multirow{2}{*}{class}  & \multirow{2}{*}{$T$}  & \multirow{2}{*}{$P$}  & \multirow{2}{*}{$S$}  &  $d_k=0$ & $d_k=1$ 
\\
 &  & & & line & point 
\\ \hline\hline
AI  & $+1$ & $\phantom{+}0$ & $0$ & $\mathbb{Z}$ 
    & 0 
    \\ \hline
BDI & $+1$ & $+1$ & $1$ & $\mathbb{Z}_2$ & $\mathbb{Z}$  
    \\ \hline
D  & $\phantom{+}0$ & $+1$ & $0$ & $\mathbb{Z}_2$ 
    & $\mathbb{Z}_2$ 
    \\ \hline
DIII & $-1$ & $+1$ & $1$ & 0 &  $\mathbb{Z}_2$ 
    \\ \hline
AII & $-1$ & $\phantom{+}0$ & $0$ & $\mathbb{Z}$ 
    & 0 
    \\ \hline
CII & $-1$ & $-1$ & $1$ & 0 & $\mathbb{Z}$ 
    \\ \hline
C  & $\phantom{+}0$ & $-1$ & $0$ & 0  
    & 0   
    \\ \hline
CI & $+1$ & $-1$ & $1$ & 0 & 0  
    \\ \hline
\end{tabular}
\hspace{0.5cm}
 \begin{tabular}{c | ccccccccccccc}
\multicolumn{4}{l}{\textbf{real case  ($d_{\mathrm{BZ}}=3$):}} \\ \hline
\multirow{2}{*}{class}   &   $d_k=0$ & $d_k=1$ & $d_k=2$
\\
 & surface  & line  & point
\\ \hline\hline
AI  & $\mathbb{Z}$ 
    & 0 & $0$ 
    \\ \hline
BDI & $\mathbb{Z}_2$ & $\mathbb{Z}$ 
    & 0 
    \\ \hline
D   & $\mathbb{Z}_2$ 
    & $\mathbb{Z}_2$ & $\mathbb{Z}$ 
    \\ \hline
DIII& 0 &  $\mathbb{Z}_2$
    &  $\mathbb{Z}_2$ 
    \\ \hline
AII & $\mathbb{Z}$ 
    & 0 & $\mathbb{Z}_2$ 
    \\ \hline
CII & 0 & $\mathbb{Z}$ 
    & 0 
    \\ \hline
C   & 0  
    & 0 & $\mathbb{Z}$  
    \\ \hline
CI  & 0 & 0  
    & 0 
    \\ \hline
\end{tabular} 
\end{center}
\caption{
  \label{single fermi}
(Symmetry of $\mathcal{H}(k)$ restricted to $S^{d_k}$)
Classification of topologically stable Fermi surfaces in two- and three-dimensional systems ($d_{\mathrm{BZ}}=2$ and $d_{\mathrm{BZ}}=3$, respectively) as a function 
of Fermi-surface dimension
$q=d_{\mathrm{BZ}}-d_k-1$ 
and symmetry class of $\mathcal{H}(k)$ restricted 
to a hypersphere $S^{d_k}$ surrounding an
individual Fermi surface.
Ten symmetry classes are distinguished, depending on the presence or absence of
 time-reversal symmetry ($T$), particle-hole symmetry ($P$), and chiral (or sublattice)
symmetry ($S$). The labels $T$, $P$, and $S$ indicate the presence or absence of time-reversal, particle-hole, and chiral symmetries, respectively, as well as the types of these symmetries.}
\end{table}

Let us construct a few simple examples of topologically stable (and unstable) Fermi surfaces
in terms of Dirac Hamiltonians defined in the continuum. 
Examples of topological Fermi surfaces defined in terms of lattice Hamiltonians will be discussed in Sec.~\ref{ss classes of gapless modes}.

\paragraph{Class A.} We first consider single-particle Hamiltonians $\mathcal{H} (k)$ with Fermi surfaces in symmetry class A, i.e., 
Fermi surfaces that are not invariant under time-reversal ($T$), particle-hole ($P$), and chiral symmetry ($S$).
Below we list examples of Hamiltonians in $d_{\mathrm{BZ}}$ spatial dimensions with Fermi surfaces in symmetry class A
\begin{eqnarray} \label{exampA}
\begin{array}{lc}
\mathrm{Hamiltonian}       &  \textrm{Fermi surface dimension $q$} 
\\ 
\mathcal{H}(k) = k_1  &  d_{BZ}-1 
\\ 
\mathcal{H}(k) =k_1 \sigma_1 + k_2 \sigma_2 &  d_{BZ}-2 
\\ 
\mathcal{H}(k) =k_1 \sigma_1 + k_2 \sigma_2 + k_3 \sigma_3 &  d_{BZ}-3
\\ 
\mathcal{H}(k) =k_1 \alpha_1 + k_2 \alpha_2 + k_3 \alpha_3 + k_4 \beta &  d_{BZ}-4
\\ 
\quad \vdots & \vdots 
\end{array}
\end{eqnarray}
Here, $\sigma_{1,2,3}$ denote the three Pauli matrices, while
$\alpha_{1,2,3}$ and $\beta$ represent the four Dirac matrices (gamma matrices).
For each example, the Fermi surface is given by the manifold $\{ k$; with $k_i = 0$ for $i=1, 2, \cdots, d_{\mathrm{BZ}} - q \}$, 
where $q$ is the dimension of the Fermi surface. 
In the above examples, Fermi surfaces with $d_{\mathrm{BZ}} -q$ odd (i.e., $d_k +1$ odd) are perturbatively stable against any
deformation of the Hamiltonian. Fermi surfaces with $d_{\mathrm{BZ}} -q$ even, on the other hand, are topologically unstable
(see Ho\v{r}ava~\cite{Horava}).
Due to the absence of a spectral symmetry (i.e., no chiral symmetry) in class A, we can add a nonzero chemical potential term $\mu \mathbbm{1}$ 
to the Hamiltonians in Eq.~(\ref{exampA}). Thus, for example, the $(d_{\mathrm{BZ}}-3)$-dimensional stable Fermi surface of
$\mathcal{H}(k) =k_1 \sigma_1 + k_2 \sigma_2 + k_3 \sigma_3$ can be turned into
a stable Fermi surface of dimension $d_{\mathrm{BZ}}-1$ upon inclusion of a finite chemical potential.
Note that the third row in the above list, i.e.,  $\mathcal{H}(k) =k_1 \sigma_1 + k_2 \sigma_2 + k_3 \sigma_3$, corresponds to a Weyl semi-metal~\cite{burkovPRL2011,burkovPRB2011,WanTurnerAschvin,PhysRevB.85.045124,PhysRevLett.107.186806,PhysRevB.86.115208}.

\paragraph{Class AIII.} Second, we consider Hamiltonians with Fermi surfaces in symmetry class AIII.
Recall that due to the presence of  chiral symmetry in class AIII (i.e., $\left\{\mathcal{H}(k), U_S \right\}=0$, where $U_S$ is an arbitrary unitary matrix), the chemical potential is pinned at $\mu=0$.
Below we list a few examples of topologically stable (and unstable) Fermi surfaces in symmetry class AIII 
\begin{eqnarray}
\begin{array}{lc}
\mathrm{Hamiltonian}       &  \textrm{Fermi surface dimension $q$}  
\\ 
\mathcal{H}(k) = k_1  \sigma_1 &   d_{BZ}-1 
\\ 
\mathcal{H}(k) =k_1 \sigma_1 + k_2 \sigma_2 &  d_{BZ}-2 
\\ 
\mathcal{H}(k) =k_1 \alpha_1 + k_2 \alpha_2 + k_3 \alpha_3 &  d_{BZ}-3
\\ 
\mathcal{H}(k) =k_1 \alpha_1 + k_2 \alpha_2 + k_3 \alpha_3 + k_4 \beta &  d_{BZ}-4
\\ 
\quad \vdots & \vdots 
\end{array}
\end{eqnarray}
Here, we find that Fermi surfaces with dimension $d_{\mathrm{BZ}} -q$ even are topologically stable, wheres
those with $d_{\mathrm{BZ}} -q$ odd are topologically unstable.

\vspace{0.5cm} 
 
In passing, we remark that the above analysis can also be applied to gapless Hamiltonians defined in an extended parameter space, i.e., Hamiltonians that are parametrized by  momentum coordinates \emph{and} some external control parameters, such as, e.g., mass terms $m_i$. The topological arguments can then be used to predict the existence of extended regions of gapless phases in the topological phase diagram
(see, e.g., Refs.~\cite{murakamiNJP07,murakami08,schnyderPRB10}).

\subsection{Comments on the stability of multiple Fermi surfaces}

It should be stressed that the above topological stability criterion (i.e., Table~\ref{single fermi})
applies only to \emph{individual} Fermi surfaces. That is, in the above analysis we considered  
the wavefunction evolution along a hypersphere $S^{d_k}$ that encloses a \emph{single} Fermi surface.
However, many lattice Hamiltonians exhibit multiple Fermi surfaces that are located
in different regions in the BZ. In that situation, one can either consider the wavefunction evolution along
hyperspheres $S^{d_k}$ that surround more than one  Fermi surface, or study the topological 
stability of each Fermi surface separately. Depending on this choice of $S^{d_k}$ one generally finds
different stability characteristics.
In the following, we make a few remarks on the topological stability of these multiple Fermi surfaces.

\paragraph{Fermion doubling.}

Due to the Fermion doubling theorem \cite{Nielsen198120},
certain topologically stable Fermi surfaces, which in the continuum limit are described in terms of Dirac Hamiltonians, (e.g., the Weyl semi-metal) cannot be realized as \emph{single} Fermi surfaces in lattice systems. That is, on a lattice  these Fermi surfaces  always 
appear in pairs with opposite K-theory charges. In that case the Fermi surfaces are not protected against commensurate perturbations, such as charge-density-wave, spin-density-wave, or other nesting-type perturbations that connect Fermi surfaces with opposite K-theory charges. 
However, these Fermi surfaces are \emph{individually} stable, i.e., stable against deformations that do not lead to nesting instabilities.

\paragraph{Effective symmetry classes.}

In the presence of multiple Fermi surfaces,  anti-unitary symmetries (i.e., TRS and PHS) can act in two different ways on
the system: (i) the symmetry maps different Fermi surfaces onto each other, or
(ii) each individual Fermi surface is (as a set) invariant under the  symmetry transformation.
In case (i) the symmetry class of the entire system is distinct from the symmetry class of each individual Fermi surface. Hence, the topological number describing the stability of an individual Fermi surface differs from the topological invariant  characterizing the entire system.

\begin{table}[t!]
\begin{center}
\begin{tabular}{c | ccc | cccccccccc}
\multicolumn{5}{l}{\textbf{complex case  ($d_{\mathrm{BZ}}=2$):}}
\\ \hline 
\multirow{2}{*}{class}  & \multirow{2}{*}{$T$} & \multirow{2}{*}{$P$}  & \multirow{2}{*}{$S$}  & $d_k=0$ & $d_k=1$ 
\\
&  & & & line & point 
 \\ \hline\hline
A & $0$ & $0$ & $0$ & $\mathbb{Z}$ & 0
\\\hline
AIII & $0$ & $0$ & $1$ & 0 & $\mathbb{Z}$ 
\\\hline
\end{tabular}
\hspace{0.5cm}
\begin{tabular}{c | ccccccccccccc}
\multicolumn{4}{l}{\textbf{complex case ($d_{\mathrm{BZ}}=3$):}}
\\ \hline
\multirow{2}{*}{class} & $d_k=0$& $d_k=1$ & $d_k=2$
\\ 
  & surface & line & point 
     \\ \hline\hline
A   & $\mathbb{Z}$ & 0
    & $\mathbb{Z}$ 
\\\hline
AIII  & 0 & $\mathbb{Z}$ & 0 
\\\hline
\end{tabular}
\end{center}
\vspace{0.5cm}
\begin{center}
\begin{tabular}{c | ccc | cccccccccc}
\multicolumn{6}{l}{
\textbf{real case  ($d_{\mathrm{BZ}}=2$):}}
\\ \hline
\multirow{2}{*}{class}  & \multirow{2}{*}{$T$} & \multirow{2}{*}{$P$}  & \multirow{2}{*}{$S$}   & $d_k=0$ & $d_k=1$ 
\\
& & &  & line & point 
\\ \hline\hline
AI & $+1$ & $\phantom{+}0$ &  $0$ & 0 & 0
    \\ \hline
BDI & $+1$  & $+1$ &  $1$ & 0 & 0 
    \\ \hline
D  & $\phantom{+}0$ & $+1$ & $0$ & $\mathbb{Z}$ 
    & 0 
    \\ \hline
DIII & $-1$ & $+1$ & $1$ & $\mathbb{Z}_2$ & $\mathbb{Z}$ 
    \\ \hline
AII & $-1$ & $\phantom{+}0$ & $0$ & $\mathbb{Z}_2$ 
    & $\mathbb{Z}_2$ 
    \\ \hline
CII & $-1$ & $-1$ & $1$ & 0 & $\mathbb{Z}_2$ 
    \\ \hline
C   & $\phantom{+}0$ & $-1$ & $0$ & $\mathbb{Z}$  & 0 
    \\ \hline
CI  & $+1$ & $-1$ & $1$ & 0 & $\mathbb{Z}$  
    \\ \hline
\end{tabular}
\hspace{0.5cm}
\begin{tabular}{c | ccccccccccccc}
\multicolumn{4}{l}{\textbf{real case  ($d_{\mathrm{BZ}}=3$):}}
\\ \hline 
\multirow{2}{*}{class} & $d_k=0$ & $d_k=1$ & $d_k=2$
\\
 & surface  & line  & point
\\ \hline\hline
AI  & 0 
    & 0 & $\mathbb{Z}$ 
    \\ \hline
BDI & 0 & 0 
    & 0 
    \\ \hline
D   & $\mathbb{Z}$ 
    & 0 & 0 
    \\ \hline
DIII& $\mathbb{Z}_2$ & $\mathbb{Z}$ 
    & 0 
    \\ \hline
AII & $\mathbb{Z}_2$ 
    & $\mathbb{Z}_2$ & $\mathbb{Z}$ 
    \\ \hline
CII & 0 & $\mathbb{Z}_2$ 
    & $\mathbb{Z}_2$ 
    \\ \hline
C   & $\mathbb{Z}$  
    & 0 & $\mathbb{Z}_2$  
    \\ \hline
CI  & 0 & $\mathbb{Z}$  
    & 0 
    \\ \hline
\end{tabular} 
\end{center}
\caption{
\label{total-table}
(Symmetry of \emph{total} system)
Alternative classification of topologically stable Fermi surfaces in
two- and three-dimensional systems ($d_{\mathrm{BZ}}=2$ and $d_{\mathrm{BZ}}=3$, respectively) as a function of
Fermi surface dimension $q=d_{\mathrm{BZ}}-d_k-1$ and
symmetry class of the \emph{total}  system.}
\end{table}

\paragraph{Classification of gapless topological phases from higher-dimensional
topological insulators and superconductors.}

Here we derive an alternative classification of gapless topological phases in terms of symmetries of the entire system (as opposed to  the symmetries of  $\mathcal{H}(k)$ restricted to a hypersphere $S^{d_k}$ surrounding an individual Fermi surface as in Sec.~\ref{topFSs}). To that end, we apply a dimensional reduction procedure to obtain $d_{\mathrm{BZ}}$-dimensional gapless topological phases from the zero-energy boundary modes of  $(d_{\mathrm{BZ}}+1)$-dimensional topological insulators (fully gapped superconductors). Namely, we observe that the surface states of $(d_{\mathrm{BZ}}+1)$-dimensional topological insulators 
can be interpreted as topologically stable Fermi surfaces in $d_{\mathrm{BZ}}$ dimensions. 
In fact, as was shown in Ref.~\cite{Gurarie},  the bulk topological invariant of a $(d_{\mathrm{BZ}}+1)$-dimensional topological insulator
 is directly related to the K-theory topological charge of the boundary Fermi surface.
Hence, we argue that the two Fermi surfaces that appear on either side of a $(d_{\mathrm{BZ}}+1)$-dimensional topological insulator can be embedded in a $d_{\mathrm{BZ}}$-dimensional BZ. Moreover, we recall from Sec.~\ref{topFSs} that the classification of stable Fermi surfaces
only depends on the codimension $(d_k+1)$ of the Fermi surface, since  a $q$-dimensional stable Fermi surface in $d_{\mathrm{BZ}}$ dimensions can always be converted into a  $(q+1)$-dimensional stable Fermi surface in $d_{\mathrm{BZ}}+1$ dimensions by including an extra momentum-space coordinate.
Based on these arguments we find that the classification of ($q=d_{\mathrm{BZ}}-d_k-1$)-dimensional Fermi surfaces in terms of symmetries of the total system is obtained from Table~\ref{periodic table of Fermi surf} with $\delta=d_k+2$, see Table \ref{total-table}.
 
Note that in the above construction of gapless topological phases 
the stable Fermi surfaces always appear in pairs (one from each of the two surfaces of the topological insulator) due to the Fermion doubling theorem. Therefore, these gapless topological phases are unstable against commensurate nesting-type deformations that connect Fermi surfaces with opposite K-theory charges.

\subsection{Bulk-boundary correspondence}
\label{bulkBndCorr}

Topological characteristics of stable Fermi surfaces  can lead to the appearance of zero-energy surface states 
via a bulk-boundary correspondence. We discuss this phenomenon in terms of a few specific examples defined in the continuum
(similar considerations can also be applied to lattice systems, cf.~Sec.~\ref{ss classes of gapless modes}).

\begin{figure}[t]
\begin{center}
\includegraphics[width=.91\textwidth]{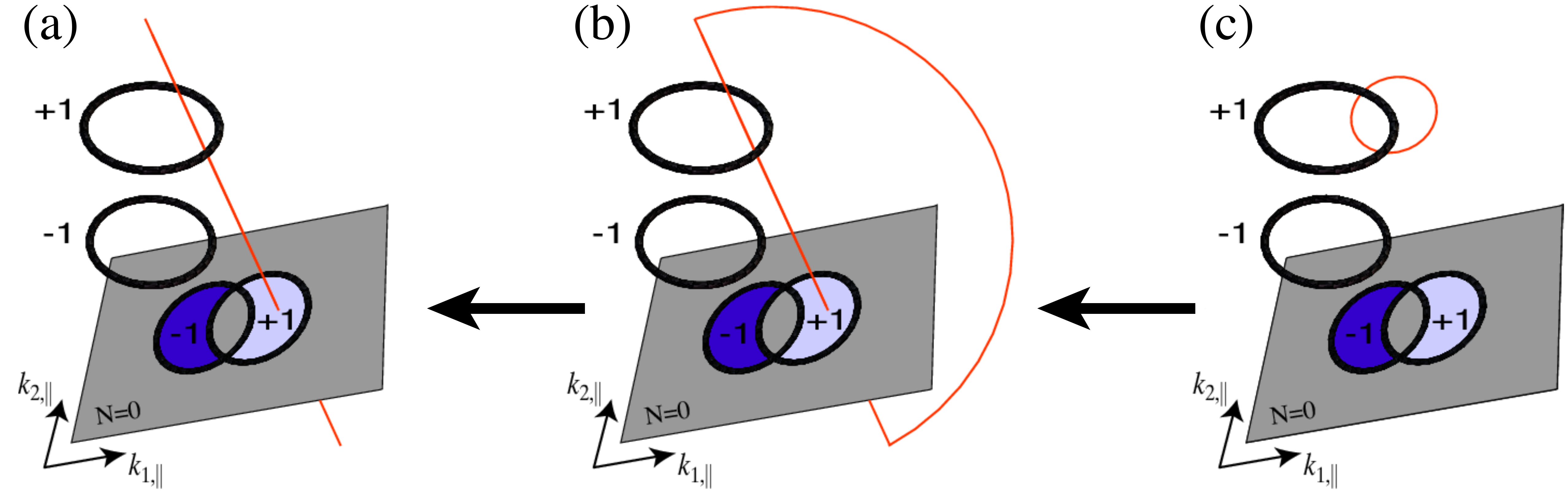} 
\caption{
Illustration of integration path deformation in the three-dimensional BZ. The gray planes represent the two-dimensional surface BZ parametrized by the two surface momenta ${k}_{1,\|}$ and ${k}_{2,\|}$. The light and dark blue areas indicate regions in the surface BZ where there appear zero-energy states.}
\label{Contour-Deform}
\end{center}
\end{figure}

\paragraph{Fermi rings in three-dimensional systems.}
Consider first the case of two topologically stable Fermi rings
in a three-dimensional system described by the Hamiltonian $\mathcal{H} ( k )$ (see Fig.~\ref{Contour-Deform}).
These rings of gapless points occur, for example, in nodal topological superconductors (e.g., class DIII, AIII, or CI, see Sec.~\ref{ss classes of gapless modes}), or in topological semi-metals with sublattice symmetry (class AIII)~\cite{paananenDahm12}.
The  topological characteristics of these Fermi rings are determined by the topology of the wavefunctions along a circle $S^{d_k=1}$ enclosing the Fermi ring (red circle in Fig.~\ref{Contour-Deform}(c)). That is, the stability of the Fermi ring is guaranteed by the conservation of a topological charge, which is given in terms of the homotopy number of the map of $S^{d_k=1}$ onto the space of Hamiltonians.

Let us now discuss the appearance of zero-energy states at a two-dimensional surface of this system. To that end, we define a two-dimensional surface BZ (gray planes in Fig.~\ref{Contour-Deform}) parametrized by the two surface momenta ${k}_{1,\|}$ and ${k}_{2,\|}$. The third momentum component, which is perpendicular to the surface BZ, is denoted by ${k}_{\perp}$. The appearance of a zero-energy state at a given
surface momentum $k^0_{\parallel} = ({k}_{1,\parallel}^{0},{k}_{2, \parallel}^{0}$)  can be understood by considering a continuous
deformation of the closed path $S^1$ in Fig.~\ref{Contour-Deform}(c) into a infinite semi-circle (Fig.~\ref{Contour-Deform}(a)), such 
that the diameter of the semi-circle is parallel to $k_\perp$ and passes through $k^0_{\parallel} = ({k}_{1,\parallel}^{0},{k}_{2, \parallel}^{0}$).
This path deformation does not alter the value of the topological number (i.e., the K-theory topological charge), as long as no Fermi ring is crossed during the deformation process. Furthermore, one can show that in the limit of an infinitely large semi-circle, the topological charge of the  Fermi ring is identical to the topological number of the one-dimensional system~$\mathcal{H} (k_\perp, k^0_{\parallel} )$.
Hence, there appear zero-energy surface states at those momenta $k^0_{\parallel}$, where the corresponding one-dimensional
gapped Hamiltonian~$\mathcal{H} (k_\perp, k^0_{\parallel} )$ has nontrivial topological characteristics, i.e., at momenta that
lie inside the projections of the Fermi rings of the bulk system. For the complex symmetry classes (i.e., class AIII for the present case) it follows that
the zero-energy surface states occur in  two-dimensional regions in the surface BZ that are bounded by the projections 
of the Fermi rings (light and dark blue areas in Fig.~\ref{Contour-Deform}) \cite{Schnyder2010,brydon2010,Schnyder2011}. In other words, the zero-energy states form two-dimensional surface flat bands.
For the real symmetry classes (i.e., for symmetries that relate $+k$ to $-k$ in momentum space), it follows that zero-energy states
 appear at certain symmetry-invariant surface momenta that lie inside the projections of the Fermi rings \cite{Schnyder2010,Schnyder2011}.

\begin{table}[t!]
 \begin{center}
  \begin{tabular}[t]{ c | c | c | c } \hline
 \multicolumn{1}{m{2.6cm}|}{\centering dimension of   BZ  ($d_{\mathrm{BZ}}$)}        &   
  \multicolumn{1}{m{2.6cm}|}{\centering  dimension of FS ($q$)}      & 
    \multicolumn{1}{m{2.6cm}|}{\centering  dim. of sphere  surrounding FS ($d_k$)}    &
  \multicolumn{1}{m{2.6cm}}{\centering  dimension of surface flat band}      \\
   \hline\hline
   2      &   0                         & 1                                 & 1             \\
   \hline
   3      &   0                         & 2                          & 1    \\
   3      &   1                         & 1                         & 2       \\
   \hline
   4      &   0                         & 3                    & 1       \\
   4      &   1                         & 2                     & 2        \\
   4      &   2                         & 1                  & 3       \\ \hline
  \end{tabular}
 \end{center}
 \caption{
Dimensionality of  zero-energy flat bands appearing at the boundary
 of a $d_{\mathrm{BZ}}$-dimensional system with a $q$-dimensional stable 
 Fermi surface in symmetry class A or AIII.}
 \label{dim of flat band}
\end{table}

\paragraph{Fermi points in three-dimensional systems.}
As a second example, let us consider a three-dimensional system $\mathcal{H} (k )$ with
two topologically stable Fermi points. Topologically stable Fermi points can be found, for example, in 
Weyl semi-metals (class A)~\cite{burkovPRL2011,burkovPRB2011,WanTurnerAschvin,PhysRevB.85.045124,PhysRevLett.107.186806,PhysRevB.86.115208}.
The stability  of these Fermi points is ensured by the conservation of the
homotopy number of the map of $S^{d_k=2}$ onto the space of Hamiltonians, where 
$S^{d_k=2}$ surrounds one of the two Fermi points. To derive the existence of surface states on a given line, say, 
$\mathcal{L} ( k^0_{2,\parallel} ) = \left\{ ( k_{1,\parallel} , k^0_{2,\parallel})  ; \,  \textrm{with } \, k_{1,\parallel} \in \mathbb{R} \, \textrm{and}  \, k^0_{2,\parallel} \, \textrm{fixed} \right\}$, within the surface BZ,
we consider a continuous deformation of the sphere $S^2$ into a half-sphere, such that the diameter of the half-sphere 
is perpendicular to the surface BZ and passes through $\mathcal{L} ( k^0_{2,\parallel} )$. 
As before, one can show that in the limit of an infinitely large half-sphere, the
topological charge of the Fermi point enclosed by the half-sphere is identical to the
topological invariant of the two-dimensional system $\mathcal{H} ( k_\perp, k_{1,\parallel} ; k^0_{2,\parallel} )$.
Thus, there appears a linearly dispersing surface state within $\mathcal{L} ( k^0_{2,\parallel} )$, whenever 
the fully gapped two-dimensional Hamiltonian
$\mathcal{H} ( k_\perp, k_{1,\parallel} ; k^0_{2,\parallel} )$ has a nontrivial topological character.
For symmetry class A, we can repeat this argument for arbitrary $k^0_{2,\parallel} $. Therefore 
 there exists a line of zero-energy modes in the surface BZ (i.e., an arc surface state) connecting the two projected 
Fermi points.

\vspace{0.5cm} 

It is straightforward to generalize the above considerations to Fermi surfaces with arbitrary codimensions,
provided that $d_k > 1$.
The result is summarized in Table.~\ref{dim of flat band}. We find that for a $q$-dimensional topologically stable Fermi surface
 in symmetry classes A or AIII, there appears a $(q+1)$-dimensional zero-energy flat band at the boundary of the system.
For $d_k =0$, however, which corresponds to stable Fermi surfaces of codimension~1 (i.e., e.g., a two-dimensional Fermi surface
in a three-dimensional BZ), there is no topological state appearing at the boundary of the system. 
The reason for this is that a Fermi surface of codimension~1 cannot
be surrounded by a hypersphere in momentum space, since the Fermi surface separates the BZ into two distinct regions.

\section{Protected surface states in nodal topological superconductors}
\label{ss classes of gapless modes}

To demonstrate the usefulness of the classification scheme of Sec.~\ref{sec:LocStab}, we study in this section
a few examples of topologically stable Fermi surfaces. Specifically,  we examine topologically stable nodal lines
in three-dimensional time-reversal invariant superconductors with and without spin-$S^z$ conservation. 
Using the generalized bulk-boundary correspondence of Sec.~\ref{bulkBndCorr}, the appearance of different types of
topological surface states is discussed.
Before introducing the specific Bogoliubov-de Gennes (BdG) model Hamiltonians in Secs.~\ref{expSzSC} and~\ref{sec:NCS}, 
we  present in Sec.~\ref{topINvs} a general derivation of  
topological invariants that characterize the stability of  nodal lines in these systems~\cite{ryuNJP10,zhangPRB08,Schnyder2010,schnyderPRB08,schnyderPRB10,volovikBOOKS,roy2008,schnyderPRL09}.
The robustness of the nodal lines and the associated  topological surfaces states
against disorder is discussed in Sec.~\ref{sec:Robust}.

\subsection{Topological invariants} 
\label{topINvs}

We start from a general 
 lattice Hamiltonian $H =  \sum_{ k } \Psi^{\dag}_k   \mathcal{H}( k ) \Psi_k $
describing time-reversal invariant 
superconductors with $N$ bands and two spin degrees of freedom.
The following derivation of $\mathbb{Z}$ topological invariants (Sec.~\ref{sec:WnO}) 
is applicable to any Hamiltonian $\mathcal{H} (k)$ with chiral symmetry, i.e., 
any $\mathcal{H} ( k)$ that anticommutes with a unitary matrix $U_S$.  This includes, in particular,
BdG Hamiltonians in symmetry class AIII, DIII, and CI, where 
chiral symmetry is realized as a combination of time-reversal and particle-hole symmetry. 
The presence of chiral symmetry implies that $\mathcal{H} (k)$ can be brought into
block off-diagonal from 
\begin{eqnarray} \label{eqChrialBasis}
\tilde{\mathcal{H}} ( k )=V\mathcal{H} (k) V^\dagger
=
\left(\begin{array}{cc}
0 & D ( k ) \\
D^{\dag} ( k ) & 0 \\
\end{array}\right),
\end{eqnarray} 
where $V$ is a unitary transformation that brings $U_S$ into diagonal form. 
In order to derive the topological invariants, it is convenient to adiabatically deform
$\tilde{\mathcal{H}}(k)$ into a flat-band Hamiltonian $Q(k)$ with eigenvalues $\pm 1$.
This adiabatic transformation does not alter the topological characteristics of $\tilde{\mathcal{H}} (k)$.
The flat-band Hamiltonian $Q(k)$ can be defined in terms of the spectral projector $P(k)$ 
\begin{eqnarray}
\label{eq:projector}
Q(k)
=
\mathbbm{1}_{4N}
-
2 P (k) 
=
\mathbbm{1}_{4N} 
- 2 \sum_{ a =1 }^{2 N}  
\left( \begin{array}{c}
\chi^-_a (k) \\
\eta^-_a (k) \\
\end{array} \right)
\left( \begin{array}{cc}
\left[ \chi^-_a (k) \right]^{\dag} &
\left[ \eta^-_a (k) \right]^{\dag} 
\end{array} \right) ,  \nonumber\\
\end{eqnarray}
where $\big( \begin{array}{c c} \chi^-_a (k) & \eta^-_a (k) \end{array} \big)^T$ are the negative-energy eigenfunctions of $\tilde{\mathcal{H}} ( k) $,
which are obtained from the eigenequation
\begin{eqnarray} \label{eigenEQ}
\left(
\begin{array}{cc}
0 & D ( k ) \\
D^{\dag} ( k ) & 0 
\end{array}
\right)
\left(
  \begin{array}{cc}
\chi^{\pm}_a ( k ) \cr
\eta^{\pm}_a ( k ) \cr
\end{array}
\right)
=
\pm \lambda_a ( k ) 
\left(
  \begin{array}{cc}
\chi^{\pm}_a ( k ) \cr
\eta^{\pm}_a ( k ) \cr
\end{array}
\right).
\end{eqnarray}
Here, 
$a=1,\ldots,2N$
denotes the combined band and spin index.
In Eq.~(\ref{eq:projector}), it is implicitly assumed that 
for the considered $k$ values
 there is a spectral gap around zero energy with
 $\left| \lambda_a ( k ) \right| > 0$, for all $a$.
By multiplying Eq.~(\ref{eigenEQ}) from the left by $\tilde{\mathcal{H}}(k)$ one can show that
the eigenfunctions of $\tilde{\mathcal{H}}(k)$ can be expressed in terms of the
eigenvectors $u_a (k)$ and $v_a (k)$ of $D (k) D^{\dag} (k) $ and $D^{\dag}(k) D(k)$, respectively,
\begin{eqnarray} \label{evalsDD}
D (k) D^{\dag} (k)  u_a (k)  = \lambda^2_a (k) u_a (k) , 
\quad
D^{\dag} (k) D (k) v_a (k)  = \lambda^2_a (k) v_a (k) ,
\end{eqnarray}
where $u_a (k)$ and $v_a(k)$ are taken to be normalized to one, i.e., $u^{\dag}_a (k) u_a (k) = v^{\dag}_a (k) v_a (k) =1$, for all $a$.
That is, we have~\cite{Schnyder2010}
\begin{eqnarray}
\label{eq:ChaiandU}
\left(
  \begin{array}{cc}
\chi^{\pm}_a ( k ) \cr
\eta^{\pm}_a ( k ) \cr
\end{array}
\right)
=
\frac{1}{\sqrt{2}} 
\left(
\begin{array}{c}
u_a (k) \\
\pm v_a (k) 
\end{array}
\right) .
\end{eqnarray}
We observe that the eigenvectors of  
$D^\dag (k)D (k)$ follow from $u_a (k)$ via 
\begin{eqnarray} \label{eq:VandU}
v_a (k) = \mathcal{N}_a (k)  D^{\dag} (k)  u_a (k) , 	 
\end{eqnarray}
with the normalization factor $\mathcal{N}_a (k) = 1 / \lambda_a (k)$.
Combining Eqs.~(\ref{eq:projector}), (\ref{eq:ChaiandU}), and (\ref{eq:VandU}) yields~\cite{Schnyder2010}
\begin{eqnarray} \label{eqQtmp}
Q(k)
=
\sum_{a=1}^{2N} \left(\begin{array}{cc}
0 & u_a (k) u^{\dag}_a (k) \frac{ D(k) }{\lambda_a (k)} \\
\frac{D^{\dag} (k) }{\lambda_a (k) } u_a (k) u_a^{\dag} (k) & 0
\end{array}\right) .
\end{eqnarray}
In other words, the off diagonal-block of $Q(k)$ is obtained as
\begin{eqnarray}
  q(k) 
  &=
  \sum_{a=1}^{2N} u_a(k)u^\dagger_a(k) 
\frac{D(k)}{\lambda_a (k)},
\quad
\textrm{where}
\quad
Q(k) &= \left(
\begin{array}{cc}
0 & q(k)\\
q^{\dag}(k)& 0
\end{array}
\right) .
\end{eqnarray}
As shown below, both  $\mathbb{Z}$ and $\mathbb{Z}_2$ topological invariants
can be conveniently expressed in terms of the unitary matrix $q(k)$.

\subsubsection{$\mathbb{Z}$ topological invariant (winding number)}
\label{sec:WnO}

Topologically stable Fermi surfaces (or nodal lines) in symmetry class AIII exist for even codimension $p=d_k+1=2n+2$ (see Table~\ref{single fermi}). The stability of these nodal lines is guaranteed by the conservation of an integer-valued topological number, namely
the winding number $\nu_{d_k=2n+1}[q]$  of $q(k)$
\begin{eqnarray} \label{eq:GenWno}
\nu_{2n+1} [q]
=
C_n
\int_{S^{2n+1}}
d^{2n+1} k \,
\epsilon^{\mu_1 \mu_2 \cdots \mu_{2n+1} }
\textrm{Tr}
\left[
q^{-1} \partial_{\mu_1} q \cdot
q^{-1} \partial_{\mu_2} q \cdots
q^{-1} \partial_{\mu_{2n+1}} q 
\right]
 ,
\nonumber\\
\end{eqnarray}
with $\epsilon^{\mu_1 \mu_2 \cdots \mu_{2n+1} }$
 the totally antisymmetric tensor and
\begin{eqnarray}
C_n 
= 
\frac{ (-1)^n n! }{ (2n+1)!} 
\left( \frac{i}{ 2 \pi } \right)^{n+1} .
\end{eqnarray}
Here, $S^{2n+1}$ denotes a hypersphere in momentum space surrounding the Fermi surface (nodal line).
The winding number $\nu_{2n+1}$ characterizes the topology of the occupied wavefunctions of $\mathcal{H}(k)$ 
restricted to $S^{2n+1}$, i.e., it describes the topology of $q(k)$ on $S^{2n+1}$. 
In other words, $\nu_{2n+1} [q]$ represents the homotopy number of the map
$S^{2n+1} \mapsto q(k) \in U (2N)$.
For $d_k=1$ (i.e., $n=0$), Eq.~(\ref{eq:GenWno}) simplifies to
\begin{eqnarray}  \label{eq1Dwno}
\nu_1 [q]
=
\frac{i}{2 \pi }
\int_{S^1} d k \,
 \mathrm{Tr} \left[  q^{-1} \partial_k q \right] 
=
-
\frac{1}{2 \pi }
\mathrm{Im}
\int_{S^1} d k \,
\mathrm{Tr} 
\left[
\partial_k  
\ln D (k)
\right] ,
\end{eqnarray}
which describes the topological stability of 
Fermi surfaces (nodal lines) of codimension $q=2$. In particular,
 $\nu_1 [q]$ 
defines the topological charge of stable nodal lines in  three-dimensional time-reversal invariant superconductors~\cite{sato06,beriPRB2010,Schnyder2010}
(see Secs.~\ref{expSzSC} and \ref{sec:NCS}).

\subsubsection{$\mathbb{Z}_2$ topological invariant}
\label{Z2invariant}

For time-reversal invariant superconductors in class DIII we can define, besides the winding number~(\ref{eq:GenWno}), 
also $\mathbb{Z}_2$ topological
numbers, provided the consider hypersphere $S^{d_k}$ surrounding the nodal line/point is left invariant under the
transformations $k \to - k$ (see Table~\ref{single fermi}).
In the following, we derive these $\mathbb{Z}_2$ numbers for the cases
$d_k=1$ and $d_k=2$, and assume that the centrosymmetric hyperspheres $S^{d_k=1}$ and $S^{d_k=2}$ contain two and four  
time-reversal invariant points $\bm{K}$, respectively. 
With these assumptions,  the $\mathbb{Z}_2$ topological numbers $W_{d_k} [q]$ can be defined in terms  
of the Paffian $\mathrm{Pf}$ of the sewing matrix $w_{a b }( k )$ 
\footnote{
The Pfaffian is an analogue of the determinant. It is defined for  $2n \times 2n$ antisymmetric matrices $A$ and can be expressed in terms of a sum
over all elements of the permutation group $S_{2n}$
\begin{eqnarray}
\mathrm{Pf} ( A) 
=
\frac{1}{2^n n! }
\sum_{\sigma \in S_{2n} }
\mathrm{sgn} ( \sigma ) \prod_{i=1}^n A_{\sigma(2i-1), \sigma ( 2 i ) } .
\nonumber
\end{eqnarray}
}, i.e. \cite{ryuNJP10,kaneMelea05a, kaneMelea05b, fuKane07, moorePRB07,royPRB09}, 
\begin{eqnarray} \label{Z2noS}
W_{d_k} [q]
&=
 \prod_{\bm{K}}
 \frac{ \mathrm{Pf}\, \left[w(\bm{K} )\right] }
 { \sqrt{ \det \left[  w ( \bm{K} ) \right] }} , \qquad \textrm{with} \quad d_k =1 ,2, 
\end{eqnarray}
where the product is over the two (four) time-reversal invariant momenta $\bm{K}$ in $S^{d_k=1}$ ($S^{d_k=2}$) and
\begin{eqnarray} \label{sewMat}
w_{a b }( k ) 
& =
\langle u^+_{a} (- k ) |\mathcal{T} \,  u^+_{b}( k ) \rangle,
\end{eqnarray}
with $a , b = 1, \ldots, 2N $. Here,  
$u^{\pm}_{a}(k)$ denotes the $a$-th eigenvector
of $Q (k )$ with eigenvalue $\pm 1$, $\mathcal{T} = \mathcal{K} \, i \sigma_2 \otimes \mathbbm{1}_{2 N}$ 
is the time-reversal symmetry operator, and $\mathcal{K}$ represents the complex conjugation operator. 
$W_{d_k} [q] = + 1 (-1)$ indicates a topologically trivial (nontrivial) character of the enclosed Fermi surface/nodal line. 
Due to the block off-diagonal structure of the flat-band Hamiltonian~(\ref{eqQtmp}),
a set of eigenvectors of $Q ( k )$, with $k \in S^{d_k}$, can be constructed as  
\begin{eqnarray} \label{EVsN}
|u^{\pm}_{{a}}( k )\rangle_{\mathrm{N}}
=
\frac{1}{\sqrt{2}}
\left(
\begin{array}{c}
n_{{a}} \\
\pm q^{\dag}( k) n_{{a}}
\end{array}
\right),
\end{eqnarray}
or, alternatively, as
\begin{eqnarray}
|u^{\pm }_{{a}}( k )\rangle_{\mathrm{S}}
=
\frac{1}{\sqrt{2}}
\left(
\begin{array}{c}
\pm  q( k ) n_{{a}} \\
n_{{a}}
\end{array}
\right),
\label{u_S}
\end{eqnarray}
where $n_{{a}}$ are $2 N$ momentum-independent orthonormal vectors. For simplicity
we choose $(n_{a})_{{b}}= \delta_{a b}$.
Observe that  both 
$|u^{\pm}_{{a}}(k)\rangle_{\mathrm{N}}$
and
$|u^{\pm}_{{a}}(k)\rangle_{\mathrm{S}}$, with $k \in S^{d_k}$,
are well-defined globally over the entire hypersphere $S^{d_k}$.
In the following we work with the basis
$|u^{\pm}_{{a}}(k )\rangle_{\mathrm{N}}$.
Eq.~(\ref{sewMat}) together with Eq.~(\ref{EVsN}) gives
\begin{eqnarray} \label{wabComp2}
w_{{a} {b} } ( k ) 
&=
\frac{1}{2} 
\big(
\begin{array}{cc}
n^{\dag}_{a}, & n^{\dag}_{a}q(- k )
\end{array}
\big)
\left(
\begin{array}{c}
q^{T}( k ) n_{b} \\
-n_{b} 
\end{array}
\right)
\nonumber \\
&=
\frac{1}{2}
\left(
n^{\dag}_{a}
q^T( k ) 
n^{\ }_{ b}
-
n^{\dag}_{ a}
q(- k ) 
n^{\ }_{b}
\right)
\nonumber \\
&=
q^T_{a b}( k ) .
\end{eqnarray}
In going from the second to the third line in Eq.~(\ref{wabComp2}), 
we used the fact that due to time-reversal symmetry
$q(- k ) = -q^T( k)$. 
Thus, the $\mathbb{Z}_2$ topological number $W_{d_k} [q]$ 
for $d_k=1$ and $d_k=2$ is
\begin{eqnarray} \label{eqW}
W_{d_k} [q]   = 
\prod_{\bm{K}} 
\frac{
\mathrm{Pf}\, \left[q^T(\bm{K})\right] 
}{
\sqrt{ \det \left[ q ( \bm{K} ) \right] }} ,
\end{eqnarray}
with $\bm{K}$ the two (four) 
 time-reversal invariant momenta of $S^{d_k=1}$ ($S^{d_k=2}$).

\subsection{Nodal topological superconductors with spin-$S^z$ conservation}

\label{expSzSC}

\begin{figure}[tb]
\begin{center}
\includegraphics[width=.95\textwidth]{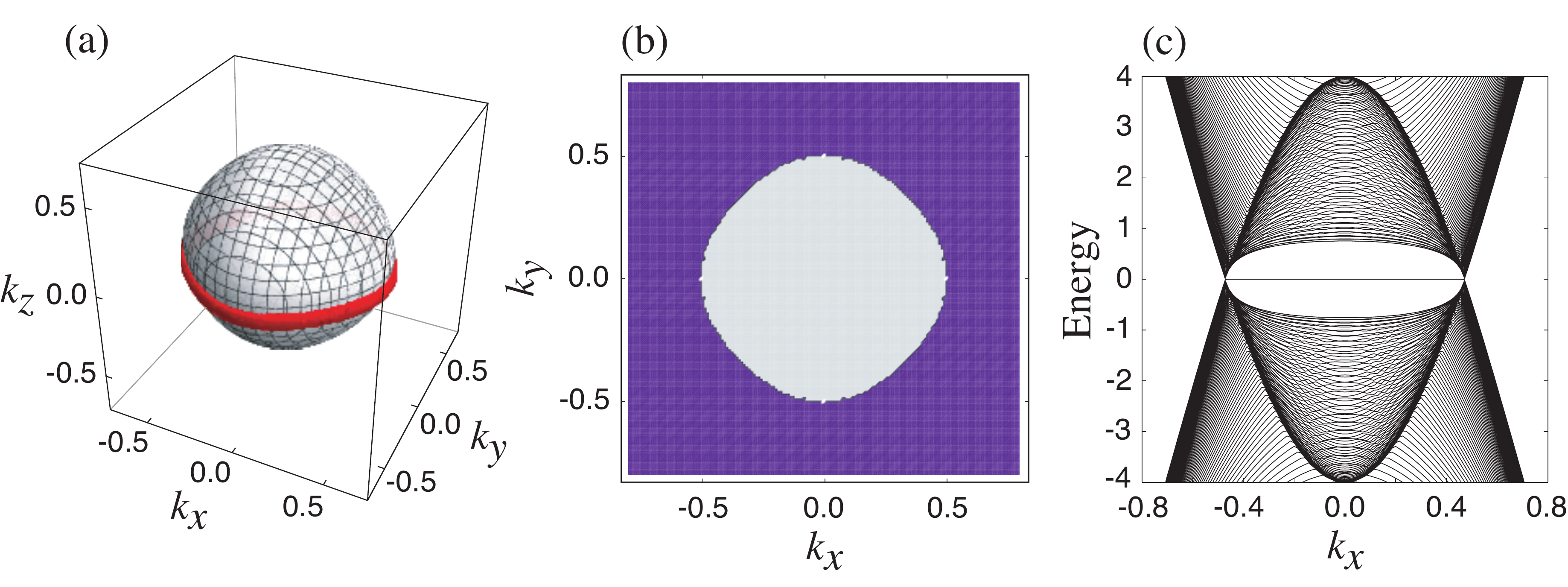}
\hspace{-0.2cm}
  \caption{
(a) Nodal structure of the energy spectrum of Hamiltonian~(\ref{AIIITBmodel}).  Here, we used the following parameters $t=2$, $\alpha=0$, $\mu=8$, $\Delta_s=0$, and $\Delta_t =2$.
(b) Winding number $\nu_1$, Eq.~(\ref{Wno12}), as a function of surface momentum for the (001) face. The color scale is such that
purple corresponds to $\nu_1=0$, whereas light gray corresponds to $\nu_1 = - 1$.
(c) Surface band structure for the (111) face as a function of $k_x$ with $k_y=0$.
}
\label{classAIIIplotEins}
\end{center}
\end{figure}

As a first  example, we study a three-dimensional time-reversal invariant superconductor with 
spin-$S^z$ conservation described by the BdG Hamiltonian 
$H = \frac{1}{2} \sum_k \psi^\dag_k \mathcal{H}_4 (k) \psi_k$, with $\psi_k =  ( c_{k\uparrow}, c_{k\downarrow}, c^{\dag}_{-k\uparrow}, c^{\dag}_{-k\downarrow}    )^{\mathrm{T}}$.
 Rotational symmetry about the $z$-axis in spin space is implemented by
$\left[ \mathcal{H}_4 ( k ), J_z \right] = 0$, with $J_z = \textrm{diag} ( \sigma_3 , - \sigma^T_3 )$. Hence, 
the $4 \times 4$  Hamiltonian $H$ can be brought into block diagonal form,
$\tilde{H} = \frac{1}{2} \sum_k \tilde{\psi}_k^{\dag} \tilde{\mathcal{H}}_4 (k) \tilde{\psi}_k$, where
$
\tilde{\mathcal{H}}_4 (k)
=
\textrm{diag} \left[
\mathcal{H}_2 ( k ) ,  - \mathcal{H}_2 ( -k) 
\right] $ and
$\tilde{\psi}_k =   ( c_{k\uparrow}, c^{\dag}_{-k\downarrow}, c^{\dag}_{-k\uparrow}, c_{k\downarrow}   )^{\mathrm{T}}$.
It follows that the topology of $\mathcal{H}_4 (k)$ is fully determined by the topology of $\mathcal{H}_2(k)$. For concreteness, 
we consider
\footnote{
 This model is equivalent to the polar state of ${}^3$He \cite{vollhardt1990superfluid}.
A two-layer version of this model might be realized in the pnictide superconductor SrPtAs \cite{Nishikubo2011,Goryo2012,2012arXiv1212.2441B}.
}
\begin{eqnarray} \label{AIIITBmodel}
\mathcal{H}_2 (k ) 
=
\left(
  \begin{array}{cc}
\varepsilon_{k} + \alpha l^z_{k} &  \Delta_s + \Delta_{t} l^z_{k}  \\
\Delta_s + \Delta_{t} l^z_{k}  &  - \varepsilon_{k} - \alpha l^z_{k}  \\
\end{array}
\right) .
\end{eqnarray} 
The normal part of this Hamiltonian, $\varepsilon_k + \alpha l^z_k = 2t\, ( \cos k_x +
\cos k_y + \cos k_z) - \mu + \alpha \sin k_z$, describes
electrons hopping between nearest-neighbor sites of a cubic lattice 
with hopping integral $t$, chemical potential $\mu$, and spin-orbit coupling strength $\alpha$. 
The superconducting order parameter contains both even-parity spin-singlet and odd-parity spin-triplet  components, 
denoted by $\Delta_s$ and $\Delta_t l^z_k = \Delta_t \sin k_z$, respectively. 
Due to the presence of time-reversal symmetry the gap functions are purely real, and hence $\mathcal{H}_2 (k)$ 
anticommutes with $\sigma_2$, i.e., $ \left\{ \mathcal{H}_2 (k), \sigma_2 \right\} = 0$.
Therefore, 
$\mathcal{H}_2 (k)$ belongs to symmetry class AIII and we find that this system can exhibit  stable nodal lines (see Table~\ref{single fermi}).
Indeed, the energy spectrum of Eq.~(\ref{AIIITBmodel}), $\lambda^{\pm}_{k} = \pm   \sqrt{ \left( \varepsilon_{k} + \alpha l^z_{k}  \right)^2 
+ \left( \Delta_s + \Delta_{t} l^z_k   \right)^2 } $, shows a nodal ring, which is located within the $(k_x,k_y)$-plane and centered around the 
$k_z$ axis (Fig.~\ref{classAIIIplotEins}(a)).
The nodal line is described by the manifold  
\begin{eqnarray} \label{nodalRing}
\left\{ k \in \textrm{BZ} \, \textrm{;  with 
$k_z = 0$ and  $k_x  = \pm  \arccos \left[   \mu / t   - 1 -   \cos k_y   \right]$} \right\} .
\end{eqnarray}
The topological stability of this nodal ring is characterized by the winding number $\nu_1$, Eq.~(\ref{eq1Dwno}).
Evaluating $\nu_1$ for Hamiltonian (\ref{AIIITBmodel}) gives
\begin{eqnarray} \label{Wno12}
\nu_1
&=&
 \frac{1}{2 \pi }
\textrm{Im} \int_{S^1} d k  \, \textrm{Tr} \left\{ \partial_{k} \ln 
\left[
  \varepsilon_{k}  -  i    \Delta_s  +  \left( \alpha  -  i   \Delta_{t} \right) l^z_{k} 
\right]
 \right\} ,
\end{eqnarray}
where $S^1$ represents a circle in momentum space. We find that $\nu = \pm 1$, whenever $S^1$ interlinks with the nodal ring (\ref{nodalRing}).
As discussed in Sec.~\ref{bulkBndCorr},  topologically nontrivial nodal lines of codimension $d_k+1=2$ in symmetry class AIII lead to
the appearance of zero-energy surface flat bands. This is demonstrated in Figs.~\ref{classAIIIplotEins}(b) and~\ref{classAIIIplotEins}(c), 
which show that zero-energy surface states appear in a two-dimensional region of the surface BZ that is bounded by the projection
of the nodal ring.

\subsection{Nodal noncentrosymmetric superconductors}
\label{sec:NCS}

As a second example, we consider nodal noncentrosymmetric superconductors.
The absence of bulk inversion symmetry in these materials leads to a spin splitting of the electronic bands
by  spin-orbit coupling. This in turn allows for the existence of mixed-parity superconducting states with
both spin-singlet and spin-triplet pairing components.
Over the past few years a number of (nodal) noncentrosymmetric superconductors have been discovered \cite{bauerSigristBook,Huang08,Klimczuk07,Bonalde2011,Lue2011,Akazawa04}, most notably 
Li$_2$Pt$_3$B \cite{yuan06,PhysRevLett.98.047002}, BiPd \cite{Bhanu2011,Mintu2012}, and the heavy-fermion
compounds CePt$_3$Si \cite{PhysRevLett.92.027003,Onuki09}, 
CeIrSi$_3$ \cite{Sugitani06}, and
CeRhSi$_3$ \cite{Kimura05}.
Recently,  nontrivial topology characteristics of nodal noncentrosymmetric superconductors have
been  studied both theoretically and experimentally 
\cite{Schnyder2010,brydon2010,Schnyder2011,satoPRB2011,yadaPRB2011,Bhanu2011,Mintu2012,PhysRevB.86.174520,2013arXiv1302.3461S,2013arXiv1302.3714B,2013arXiv1302.1943T}.
Specifically, it was found that noncentrosymmetric superconductors belong to symmetry class DIII,
which, according to Table \ref{single fermi}, implies that  three-dimensional  noncentrosymmetric superconductors
can support topologically stable nodal lines.
To exemplify the topological features of these nodal
superconductors we study in this subsection a simple BdG model Hamiltonian 
describing a single-band nodal noncentrosymmetric superconductor with monoclinic crystal symmetry $C_2$ (relevant for BiPd).
Implications of some of our findings for experiments on BiPd will be discussed at the end of this subsection.

\paragraph{Model definition.} 
We start from the BdG Hamiltonian
$
H= \frac{1}{2} \sum_k \psi^{\dag}_k  \mathcal{H}(k) \psi_k,
$
where $ \psi_k = ( c_{k \uparrow} , c_{k \downarrow} , c^\dagger_{-k \uparrow}, c^\dagger_{-k \downarrow} )^{\mathrm{T}}$ and
\begin{eqnarray} \label{defHamNCS}
\mathcal{H}(k)=\left(\begin{array}{cc}h(k) & \Delta(k) \\\Delta^\dagger(k) & -h^T(-k)\end{array}\right).
\end{eqnarray} 
The normal state Hamiltonian 
\begin{eqnarray} \label{NCS_normSt}
h(k)&=\epsilon_k \mathbbm{1}_{2\times 2}+\alpha \, \vec{l}_k \cdot \vec{\sigma} + \vec{B} \cdot \vec{\sigma}, 
\end{eqnarray}
describes electrons hopping on a cubic lattice with dispersion $\epsilon_k=2 t(\cos k_x+\cos k_y+\cos k_z)-\mu$ and
Rashba-type antisymmetric spin-orbit coupling $\alpha \, \vec{l}_k \cdot \vec{\sigma}$.
Here, $t$ denotes the hopping amplitude, $\mu$ the chemical potential, $\alpha$ the spin-orbit coupling strength, and
$\vec{l}_k = - \vec{l}_{-k}$ the antisymmetric spin-orbit coupling potential. $\vec{\sigma}=(\sigma_1,\sigma_2,\sigma_3)^{\mathrm{T}}$ represents the vector of Pauli matrices. We have included in Eq.~(\ref{NCS_normSt})  a Zeeman term  $\vec{B} \cdot \vec{\sigma}$, which  allows us to 
study the stability of the topological surface states against time-reversal symmetry breaking perturbations.
The superconducting order parameter $\Delta (k)$ is in general an admixture of
even-parity spin-singlet $\Delta_s$ and odd-parity spin-triplet $\Delta_t \vec{d}_k$ components
\begin{eqnarray} \label{SCpartNCS}
\Delta(k)&= \left( \Delta_s \mathbbm{1}_{2\times 2} + \Delta_t \,  \vec{d}_k \cdot \vec{\sigma} \right) \left( i \sigma_y \right) .
\end{eqnarray}
In the following we assume that the
spin-triplet pairing vector $\vec{d}_{k}$ is aligned with the spin-orbit coupling vector $\vec{l}_k$, i.e., 
we set $\vec{d}_{k} = \vec{l}_k$.
For the numerical computations we will set 
$(t,  \mu, \alpha, \Delta_t) =
( -0.5, -2.0,1.0,1.0 )$.
The particular form of $\vec{l}_k$ is constrained by the lattice symmetries of the noncentrosymmetric crystal \cite{samokhinAnnals09}.
Within a tight-binding expansion, we obtain for the monoclinic crystal point group $C_2$ to lowest order 
\begin{eqnarray} \label{defLvec}
\vec{l}_k =\left(\begin{array}{c}a_1\sin k_x +a_5\sin k_y \\a_2\sin k_y +a_4\sin k_x \\a_3\sin k_z\end{array}\right),
\end{eqnarray}
where $a_{i}$ ($i=1,\ldots, 5$) are five real parameters.

With the above parametrization, Hamiltonian~(\ref{defHamNCS}) in the absence of a Zeeman magnetic field, i.e.\ $\vec{B}=0$,  generically
exhibits stable nodal lines in the three-dimensional BZ.
These nodal lines are in symmetry class DIII (or class AIII if the hypersphere $S^{d_k}$ surrounding the nodal lines is not centrosymmetric), since Hamiltonian~(\ref{defHamNCS})  is invariant under both time-reversal symmetry $\mathcal{T}=U_T\mathcal{K}$, with $\mathcal{T}^2 = -1$, and particle-hole symmetry $\mathcal{P}=U_P\mathcal{K}$, with $\mathcal{P}^2=+1$.
Time-reversal symmetry $\mathcal{T}$ acts on the BdG Hamiltonian as
\begin{eqnarray}
U_T^{-1}\mathcal{H}(k)U_T =\mathcal{H}^*(-k),
\end{eqnarray}
where $U_T= \mathbbm{1}_{2\times 2}\otimes i \sigma_2$.
Particle-hole symmetry $\mathcal{P}$
is implemented as
\begin{eqnarray}
U_P^{-1}\mathcal{H}(k)U_P =-\mathcal{H}^*(-k),
\end{eqnarray}
where $U_P=\sigma_1 \otimes \mathbbm{1}_{2\times 2}$.
Observe that upon inclusion of a finite Zeeman magnetic field, $\vec{B} \ne 0$,
time-reversal symmetry is broken, whereas particle-hole symmetry
remains satisfied. I.e., Hamiltonian (\ref{defHamNCS}) restricted to $S^{d_k}$ with $\vec{B} \ne 0$ belongs to symmetry class~D
(or class A if $S^{d_k}$ is not centrosymmetric) 
and can therefore no longer support stable nodal lines (see Table~\ref{single fermi}).

\paragraph{Nodal structures.}

\begin{figure}[t!]
\center
\subfigure{
\includegraphics[height=3.5cm]{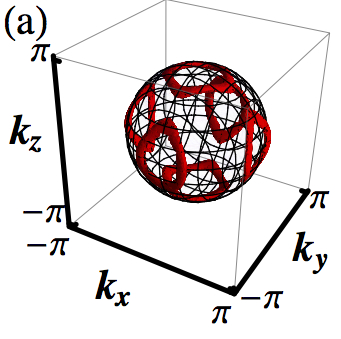}
\label{nodal1}
}
\subfigure{
\includegraphics[height=3.5cm]{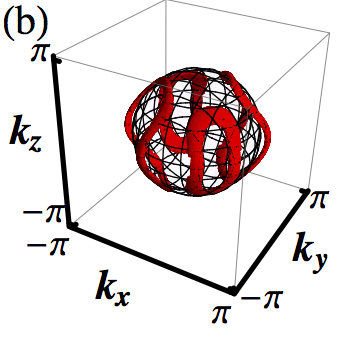}
\label{nodal2}
}
\subfigure{
\includegraphics[height=3.5cm]{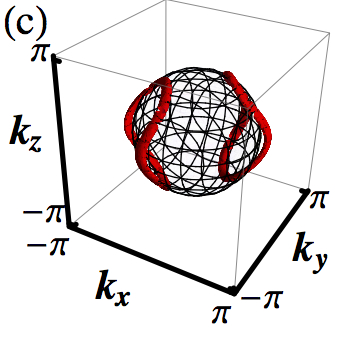}
\label{nodal3}
}
\subfigure{
\includegraphics[height=3.5cm]{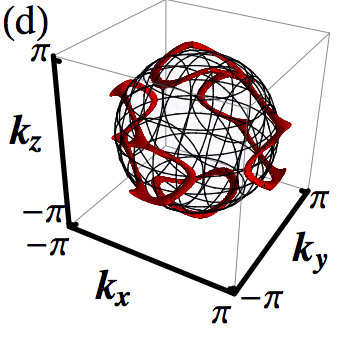}
\label{nodal4}
}
\subfigure{
\includegraphics[height=3.5cm]{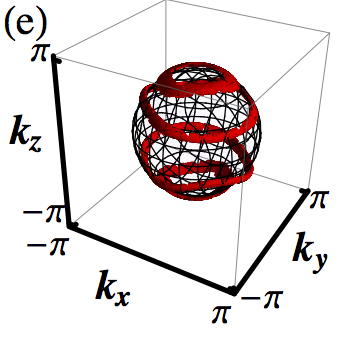}
\label{nodal5}
}
\subfigure{
\includegraphics[height=3.5cm]{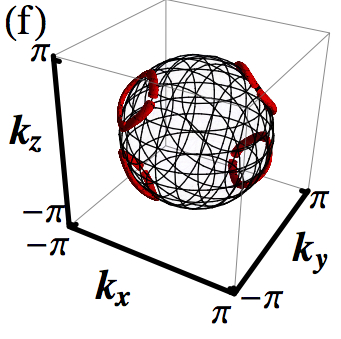}
\label{nodal6}
}
\subfigure{
\includegraphics[height=3.5cm]{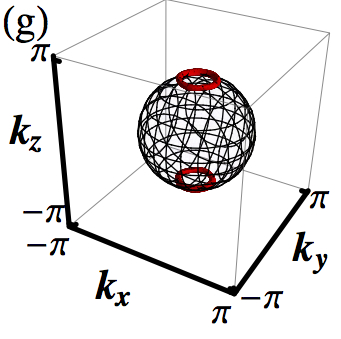}
}
\caption{
  \label{nodal}
Nodal structure of $\lambda^-_{k} = \sqrt{  ( \xi^-_k  )^2 +  ( \Delta^-_k  )^2 }$ for different parameter choices 
$(\Delta_s, a_1=a_2, a_3, a_4=a_5)$:
(a) $(1.0, 1.0, 0.5, 0.0)$,
(b) $(1.0, 0.9, 0.0, 0.1)$, 
(c) $(1.0, 0.5, 0.0, 0.5)$, 
(d) $(1.5, 0.7, 1.8, -0.7)$, 
(e) $(1.0, 0.2, 1.1, -0.2)$,  
(f) $(1.8, 0.6, 1.3, 0.6)$,
and (g) $(0.6, 1.0, 0.0, 0.0)$.  
The transparent surface represents the negative helicity Fermi surface given by $\xi^-_k = 0$.}
\end{figure}

The energy spectrum of Eq.~(\ref{defHamNCS}) with $\vec{B}=0$ is given by
$\left\{   - \lambda^-_{1, k } ,  - \lambda^+_{2,k } , + \lambda^-_{1, k } ,  + \lambda^+_{2,k }  \right\}$ with
$\lambda^-_{k} = \sqrt{  ( \xi^-_k  )^2 +  ( \Delta^-_k  )^2 }$ 
and
$\lambda^+_{k} = \sqrt{  ( \xi^+_k  )^2 +  ( \Delta^+_k  )^2 }$.
Here, we have introduced the shorthand notation $\xi^{\pm}_k = \varepsilon_k \pm \alpha  | \vec{l}_k  | $
and $\Delta^{\pm}_k = \Delta_s \pm \Delta_t   | \vec{l}_k  | $.
Without loss of generality  we can take $\Delta_s$, $\Delta_t >0$, in which case the positive helicity band $\lambda^+_k$ is always fully gapped,
whereas the negative helicity band $\lambda^-_k$  exhibits nodal lines.
In Fig.~\ref{nodal} we study the nodal structure of the negative helicity band as a function of
$\Delta_s $,
$a_1 = a_2$, $a_3$, and
$a_4 = a_5$.
The topological stability of these nodal rings is guaranteed by the winding number $\nu_1$, Eq.~(\ref{eq1Dwno}).
Using Eqs.~(\ref{defHamNCS}), (\ref{NCS_normSt}), and (\ref{SCpartNCS}) we find that %
\footnote{In certain cases, provided that $S^1$ is centrosymmetric, the stability of the nodal lines
is also protected by the $\mathbb{Z}_2$ number $W_1$, Eq.~(\ref{eqW}), see Refs.~\cite{Schnyder2010,Schnyder2011}.}
\begin{eqnarray} \label{nuNCSex}
\nu_1
&=&
 \frac{1}{2 \pi }
\textrm{Im} \int_{S^1} d k  \, \textrm{Tr} \left\{ \partial_{k} \ln 
\left[
 \left( \varepsilon_k + i \Delta_s \right) \mathbbm{1}_{2 \times 2}
 + \left( \alpha + i \Delta_t \right)   \vec{l}_k \cdot \vec{\sigma}  
\right]
 \right\} ,
\end{eqnarray}
where $S^1$ is a circle in momentum space.
For brevity, we discuss only two parameter choices, namely $(\Delta_s, a_1, a_3, a_4) = (1.0, 0.5, 0.0, 0.5)$ 
(Fig.\ \ref{nodal3}) and $(1.8, 0.6, 1.3, 0.6)$ (Fig.~\ref{nodal6}), which we refer
to as  ``Case-1" and ``Case-2", respectively.
To determine the topological charges $\nu_1$ of the nodal rings for these two cases we consider a noncontractible circle $S^1$ along the (100) direction of the BZ torus $T^3$. 
We find that for Case-1 the nodal ring located
within the half-space $k_y >0$ ($k_y<0$) of Fig.\ \ref{nodal3} carries
 topological charge $\nu_1=+1$ ($-1$). For Case-2 the nodal ring in the first and fifth octants of the BZ with $(\mathop{\textrm{sgn}} k_x , \mathop{\textrm{sgn}} k_y, \mathop{\textrm{sgn}} k_z ) = (+,+,+)$ and $(+,+,-)$, respectively, have $\nu_1 = +1$, whereas the rings in the third and seventh octants with  
$(\mathop{\textrm{sgn}} k_x , \mathop{\textrm{sgn}} k_y, \mathop{\textrm{sgn}} k_z ) = (-,-,+)$ and $(-,-,-)$, respectively, have $\nu_1 =-1$.

\begin{figure}[t!]
\center
\subfigure{
\label{2ringsflatband}
\includegraphics[height=3.8cm]{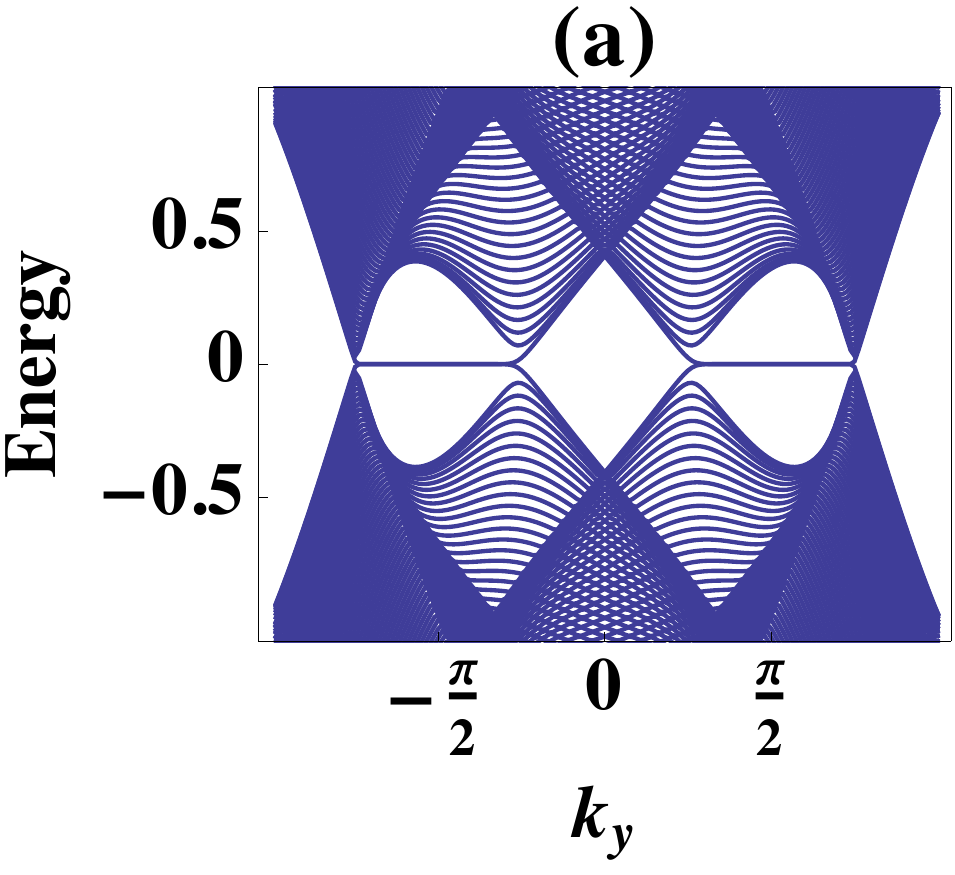}
}
\subfigure{
\label{2ringsflatbandz05}
\includegraphics[height=3.8cm]{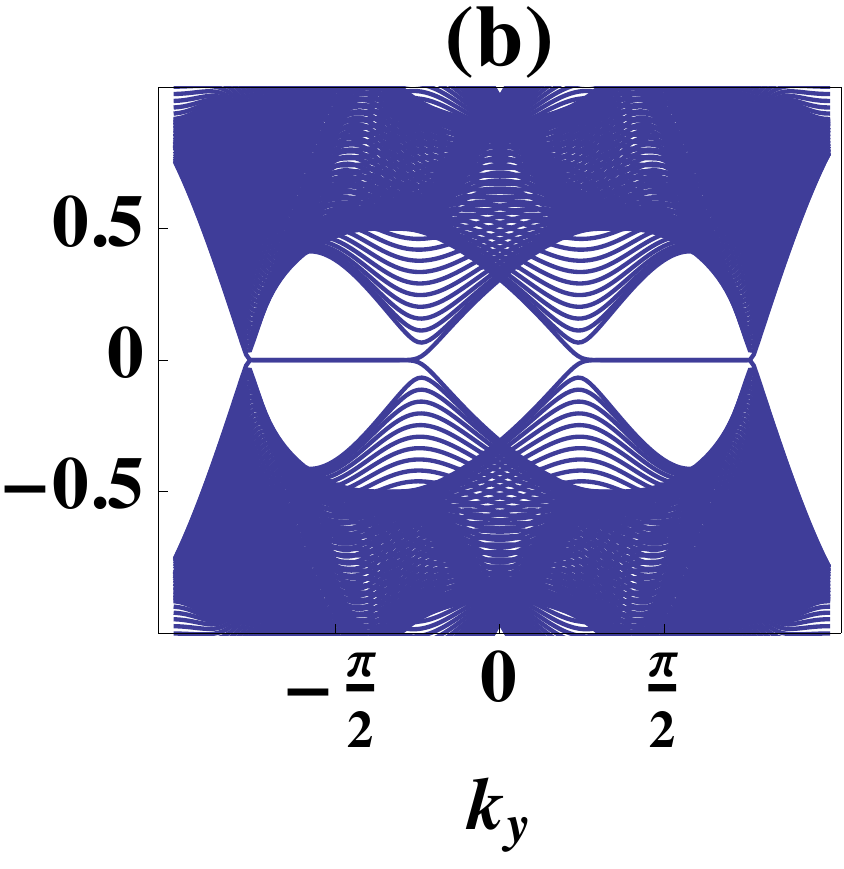}
}
\subfigure{
\label{2ringsflatbandx03}
\includegraphics[height=3.8cm]{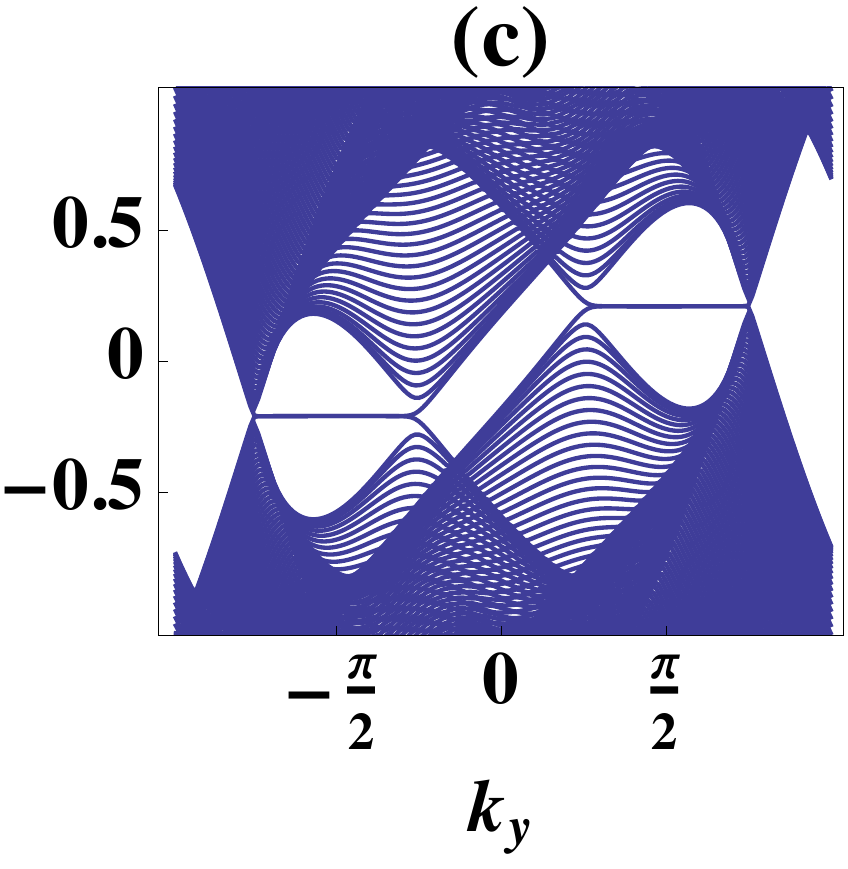}
}
\subfigure{
\label{4ringsflatband}
\includegraphics[height=3.8cm]{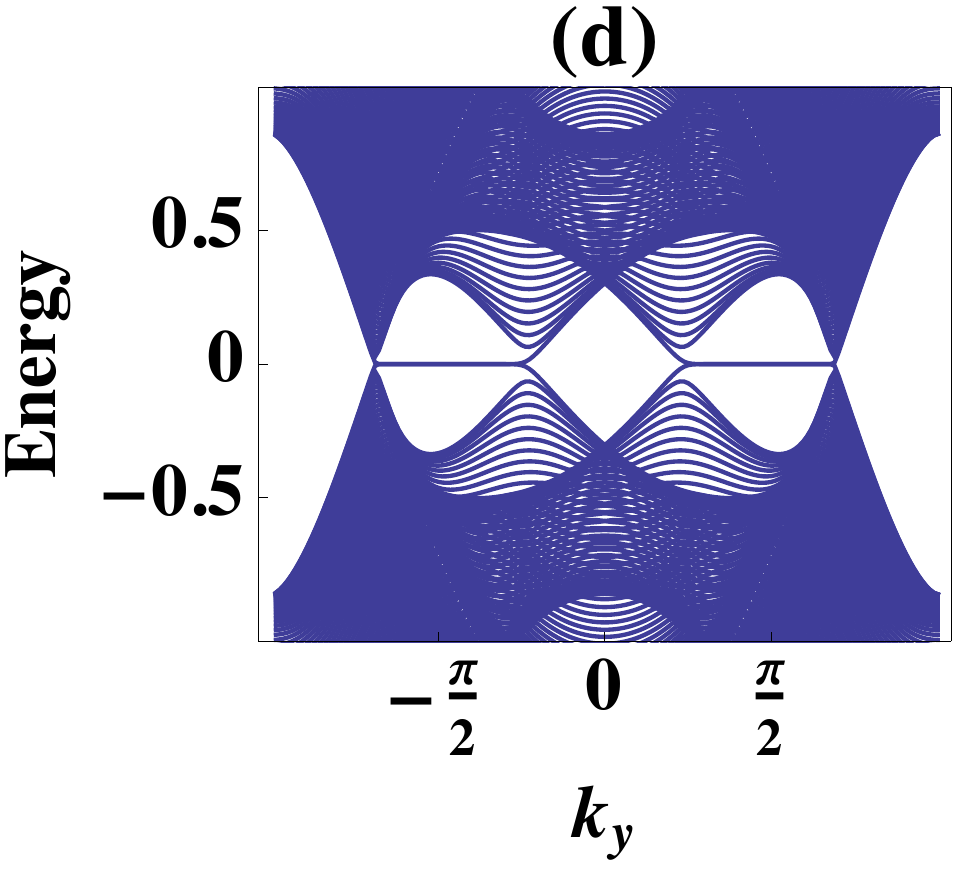}
}
\subfigure{
\label{4ringsflatband_z}
\includegraphics[height=3.8cm]{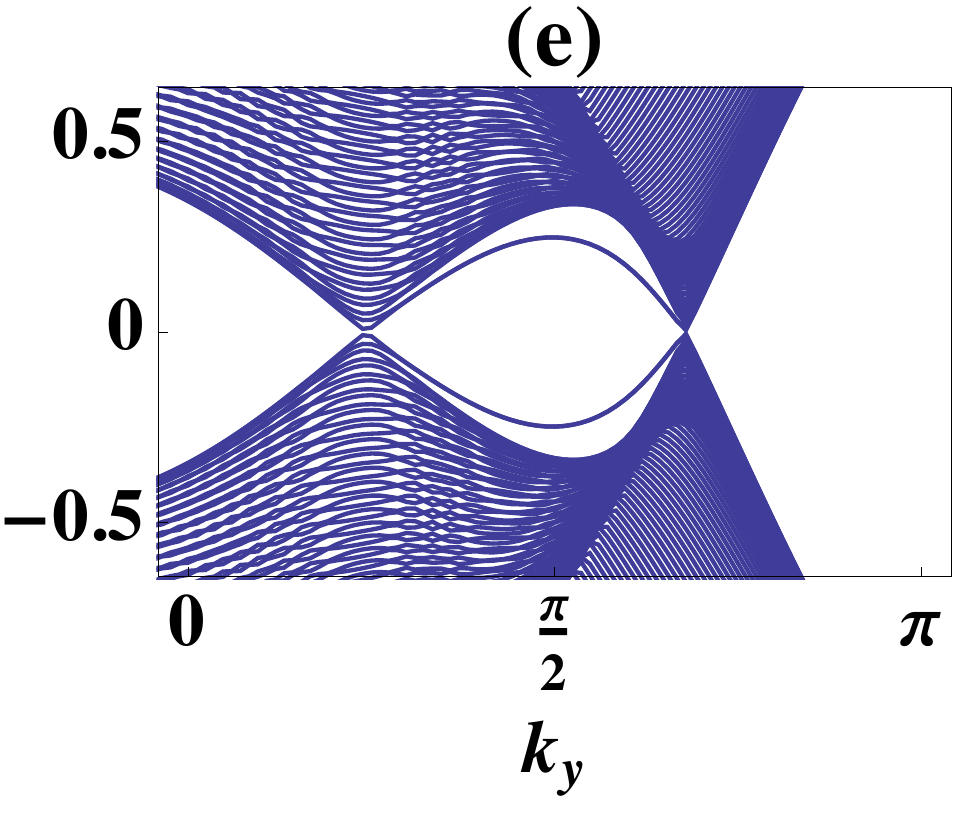}
}
\caption{ \label{fig4}
Surface band structure of 
the noncentrosymmetric superconductor (\ref{defHamNCS}).
(a)-(c): Band structure for the parameter choice Case-1 on the $(100)$ face
as a function of surface momentum $k_y$ with $k_z=0$ and
(a) $\vec{B}=(0,0,0)$, (b) $\vec{B}=(0,0,0.5)$, and (c) $\vec{B}=(0.3,0,0)$.
(d) and (e):  
Band structure for parameter choice Case-2  in the absence of a Zeeman magnetic field
on (d) the (100) face with $k_z=\pi/2$,
and (e)  the (001) face with $k_x=\pi/2$.}
\label{surface}
\end{figure}

\setcounter{footnote}{0}

\paragraph{Surface band structure.} \label{paraSurfBand}  
Due to the bulk-boundary correspondence of Sec.~\ref{bulkBndCorr},  
the topologically nontrivial nodal lines of Hamiltonian~(\ref{defHamNCS})
give rise to zero-energy surface flat bands. That is, zero-energy states appear
within two-dimensional regions of the surface BZ that are bounded by the projected bulk nodal lines.
This is illustrated in Fig.~\ref{fig4}, which shows the surface band structure of Hamiltonian~(\ref{defHamNCS})
on the (100) and (001) faces. On the (100) face, the projected nodal rings for both Case-1 and \mbox{Case-2} do not
overlap leading to several zero-energy surface flat bands (see Figs.\  \ref{2ringsflatband}
 and  \ref{4ringsflatband}). Similar considerations also hold for the (010) surface, since for the considered parameter choices
the absolute value of the spin-orbit coupling vector
 $| \vec{l}_k  |$ is symmetric under the interchange of $k_x$ with $k_y$.
 On the (001) face for Case-2, on the other hand, two projected nodal rings with opposite topological
 charge overlap, and hence the two topological charges cancel. Consequently, there are no flat-band states appearing
 on the (001) face for Case-2 (see Fig.~\ref{4ringsflatband_z})
 \footnote{Recall that the topological charge of a given nodal ring depends on the chosen integration path $S^1$ in Eq.~(\ref{nuNCSex}).
 Consequently, the topological charge of a \emph{projected} nodal ring in the surface BZ  depends on the surface orientation.}. 
 This is in fact a generic property of model~(\ref{defHamNCS}). Since  
 $| \vec{l}_k |$ is
 symmetric under $k_z \to - k_z$, the topological charges of the projected nodal rings in the surface BZ of the (001) face always add up to zero. Hence, there are no zero-energy flat bands appearing on the (001) surface, irrespective of the parameter choice for $\vec{l}_k$, Eq.~(\ref{defLvec}).

In Figs.~\ref{2ringsflatbandz05} and \ref{2ringsflatbandx03}, we study how the surface band structure on the (100) face for Case-1 is modified
in the presence of a time-reversal symmetry breaking Zeeman field $\vec{B} \ne 0$. 
Interestingly, we find that  a field along the $z$-axis leaves the flat bands  unaffected (Fig.~\ref{2ringsflatbandz05}), whereas a field within the $x-y$ plane gives rise to an energy shift of the flat-band states (Figs.~\ref{2ringsflatbandx03}). This behavior can be explained in terms of the strong spin polarization of the flat-band states. It turns out that the surface flat bands of Fig.~\ref{2ringsflatband} are spin polarized within the $x-y$  plane, and consequently  a  field along the $z$-axis does not couple to them.

\begin{figure}[t!]
\center
\subfigure {
\label{2ringsLDOS}
\includegraphics[height=3.5cm]{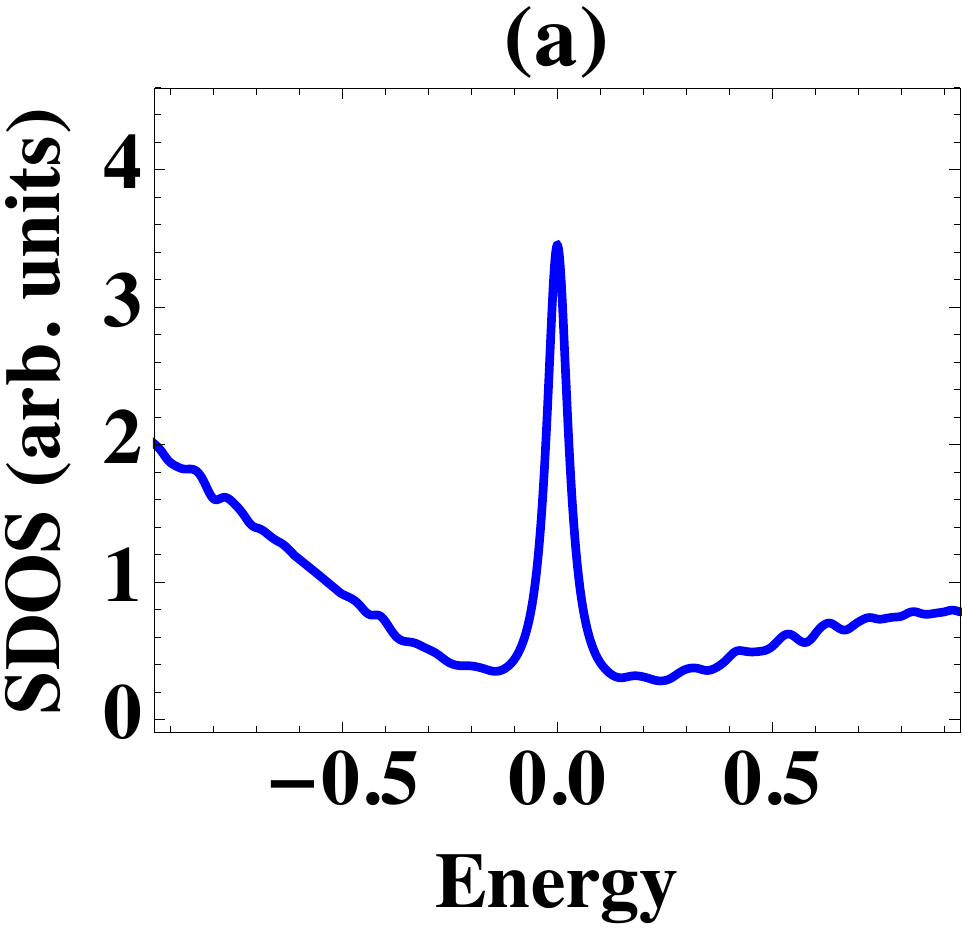}
}
\subfigure {
\label{2ringsLDOSz03}
\includegraphics[height=3.5cm]{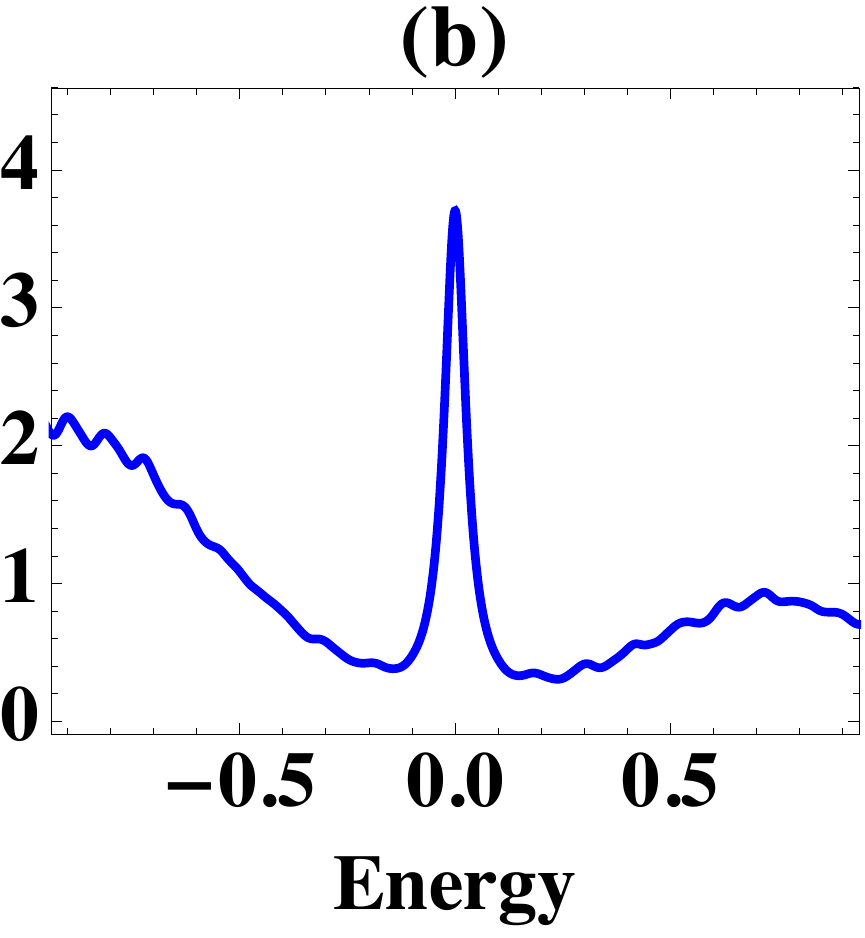}
}
\subfigure{
\label{2ringsLDOSz05}
\includegraphics[height=3.5cm]{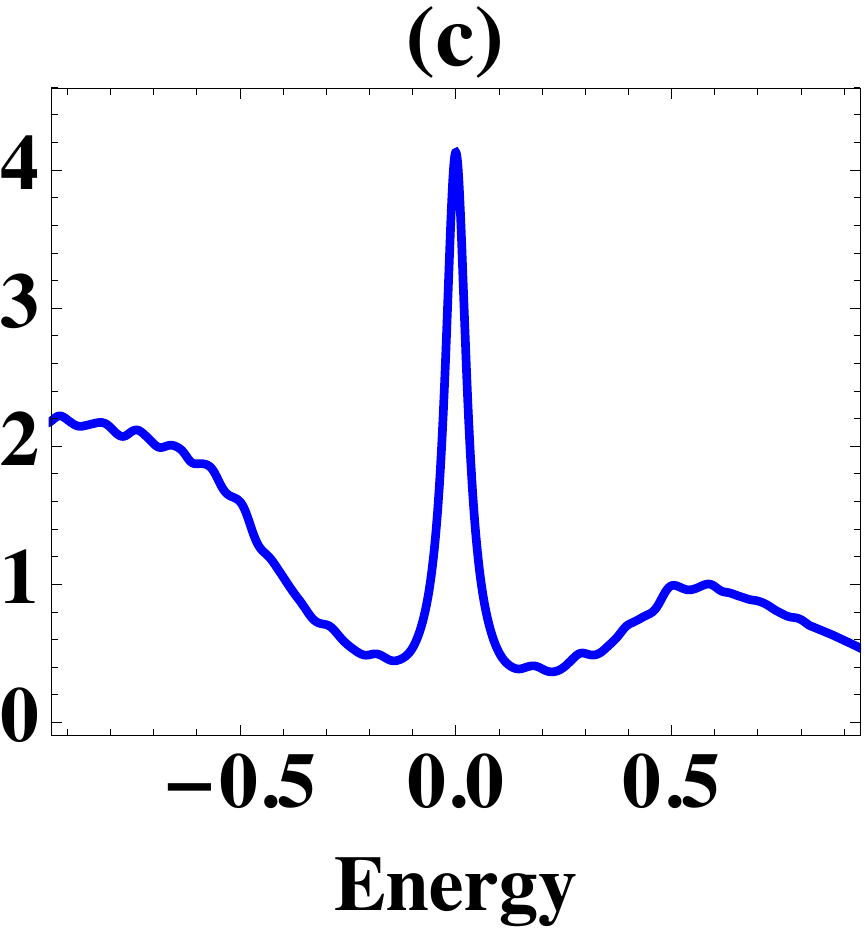}
}
\subfigure {
\label{2ringsLDOSx03}
\includegraphics[height=3.5cm]{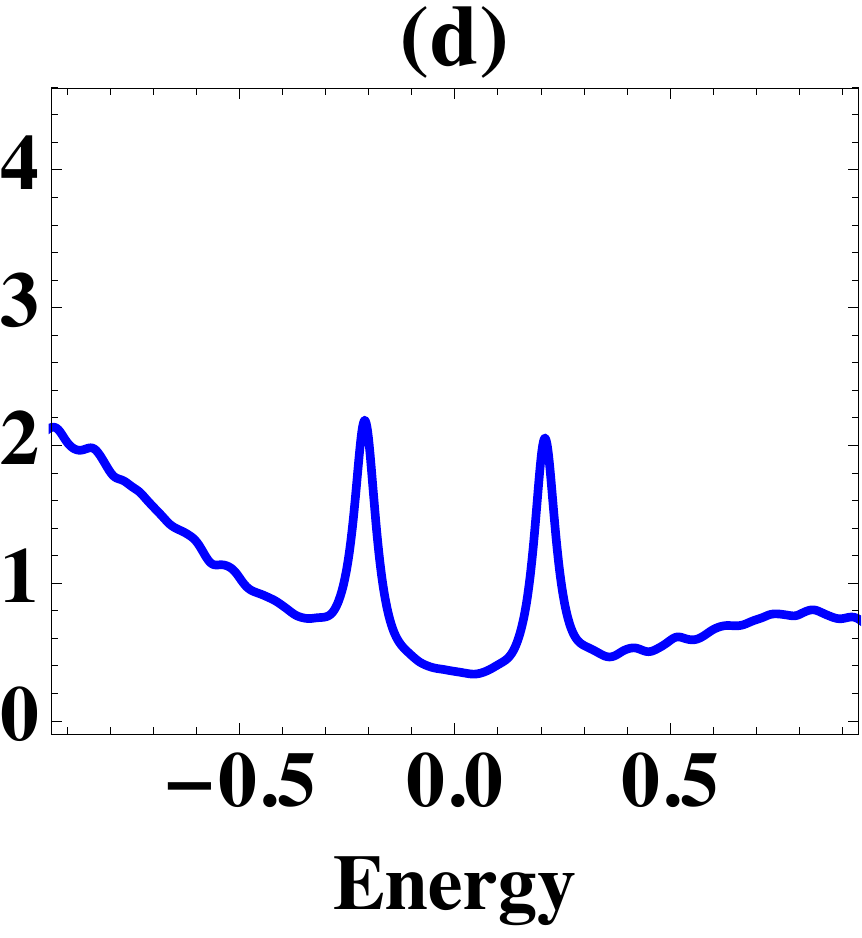}
}
\caption{
Surface density of states $D (x=1, E)$, Eq.~(\ref{formSDOS}), at the (100) face of the noncentrosymmetric superconductor (\ref{defHamNCS})
with parameter choice Case-1 (a) in the absence of a  Zeeman magnetic field,
(b) for $\vec{B}=(0, 0, 0.3)$, (c) for $\vec{B}=(0, 0, 0.5)$, and  (d) for $\vec{B}=(0.3, 0, 0)$.}
\label{LDOS1}
\end{figure}

\paragraph{Surface density of states.}
Surface flat bands manifest themselves as a zero-energy divergence in the surface density of states, and hence give rise to
a zero-bias peak in the tunneling conductance \cite{brydon2010,Schnyder2011}. This zero-bias conductance peak 
depends strongly on the surface orientation, due to the changing projection of the bulk nodal rings onto the surface BZ. 
To illustrate this dependence, let us compute the surface density of states  of the noncentrosymmetric superconductor
(\ref{defHamNCS}). The density of states in the $x$-th layer from, e.g., the (100) surface is given by
\begin{eqnarray} \label{formSDOS}
&  D(x,E)=
  \frac{1}{N_yN_z}\sum_{n, k_\parallel }
\Big[
  (|v_{n \uparrow}(x,k_\parallel )|^2+|v_{n \downarrow}(x, k_\parallel )|^2 )
  \delta(E- E_{n} ( k_\parallel )) 
  \nonumber \\ 
&  + (|u_{n \uparrow}(x, k_\parallel )|^2+|u_{n \downarrow}(x,k_\parallel )|^2 )
  \delta(E+ E_n ( k_\parallel )) \Big],
\end{eqnarray}
where $k_{\parallel} = (k_y,k_z) $ represents the surface momenta,
$N_y$  and $N_z$ are the number of $k_y$ and $k_z$ points, respectively,  in the surface BZ, 
and
$\phi_{n} (x, k_\parallel) = (u_{n \uparrow}(x, k_\parallel ), u_{n \downarrow}(x, k_\parallel),v_{n \uparrow}(x, k_\parallel ), 
v_{n \downarrow}(k_\parallel))$ denotes the eigenvector  of  
$\mathcal{H}(x,x',k_\parallel) = \frac{1}{2 \pi } \int d k_x \, e^{i k_x ( x - x' )} \mathcal{H} (k )  $ 
with eigenenergy $E_n ( k_\parallel )$, i.e., $\mathcal{H}(x,x',k_\parallel) \phi_n ( x', k_{\parallel} ) = E_n ( k_{\parallel} ) \phi_n (x, k_{\parallel} )$. 
Fig.~\ref{LDOS1} displays the surface density of states $D(x=1, E)$ at the (100) face of $\mathcal{H} (k)$ for the parameter choice Case-1, both in the presence and in the absence of an external magnetic field $\vec{B}$. 
While a field along the $z$-axis does not split the zero-energy peak (Figs.~\ref{2ringsLDOSz03} and (c)), we find that a field with a finite component
in the $x-y$ plane leads to a splitting which is roughly proportional to the field strength $ | \vec{B}  |$ (Fig.~\ref{2ringsLDOSx03}, cf.\ also Figs.~\ref{2ringsflatbandx03}).

\paragraph{Implications for experiments on BiPd.}
BiPd is a  noncentrosymmetric superconductor with transition temperature T$_C$ = 3.8 K \cite{Bhanu2011}.
Recently, a zero-bias conductance peak in point contact measurements of BiPd single crystals has been reported \cite{Mintu2012}.
Due to the large number of atoms in its unit cell, BiPd has low crystal symmetry, with monoclinic point group $C_2$ and no center of inversion. Moreover, bismuth has a high atomic number, which gives rise to strong spin-orbit coupling. These two properties together likely lead to
multiple spin-split Fermi surfaces and, correspondingly, to a superconducting state with multigap character.
It is not unlikely that at least one of these multiple gaps has  nontrivial topological properties. 

Even though the simple single-band Hamiltonian (\ref{defHamNCS}) shares the same point group symmetry $C_2$ with BiPd, it only provides a very crude model for this noncentrosymmetric superconductor.  An accurate phenomenological description of BiPd would require detailed knowledge of the band structure,  the pairing symmetry, and the pairing  mechanism of this superconductor.  In the absence of this information, we can only speculate about possible origins of the zero-bias conductance peak observed in BiPd. According to Ref.~\cite{Mintu2012}, one plausible explanation is that BiPd is a nodal topological superconductor with zero-energy surface flat bands. A pronounced zero-bias conductance peak has been observed for contacts both on  the (010) face and on the (001) face, although in the latter case the intensity is somewhat weaker \cite{Mintu2012}.
While the  zero-bias peak for the (010) surface is in agreement with  Hamiltonian~(\ref{defHamNCS}), the conductance peak for the (001) face cannot be explained within this simple model description (see discussion on page \pageref{paraSurfBand}). 
The source of this discrepancy probably lies in the oversimplified assumptions made for the band structure and gap functions
entering in model (\ref{defHamNCS}). Impurity scattering and interface roughness are further complications that need to be taken into account
when interpreting the interesting point contact conductance measurements of Ref.~\cite{Mintu2012}.

\subsection{Robustness of  surface flat bands against disorder}
\label{sec:Robust}

Determining the robustness of  gapless topological phases and their topological surface states against disorder 
requires a careful analysis of different types of scattering processes involving both quasiparticle bulk and surface states. 
How susceptible any topological Fermi surface is to impurity effects crucially depends on symmetry properties and on  their codimension~$p=d_k+1$.
The latter follows, for instance, from a simple renormalization group argument which shows that uncorrelated (or short-range) correlated on-site disorder
is marginal by power counting for Fermi surfaces with $p = 2$ and irrelevant for Fermi surfaces with $p>2$.
Since a detailed analysis of disorder effects in gapless topological phases is beyond the scope of this paper, we focus here mostly
on impurity effects in nodal noncentrosymmetric superconductors (cf.\ Sec.~\ref{sec:NCS}), and only briefly discuss
some general arguments suggesting that the surface flat bands of these systems are partly protected against disorder.

\textit{(i)} 
First, we observe that similar to other topological systems  with strong spin-orbit interactions \cite{hasanKaneReview10,qiZhangReview10,HasanMooreReview11,ryuNJP10}, the  surface states
of noncentrosymmetric  superconductors possess a helical spin texture \cite{2013arXiv1302.3461S,2013arXiv1302.3714B}. That is,
the surface flat bands are strongly spin polarized, with states with opposite momenta exhibiting opposite spin polarization.  This nontrivial spin texture leads to the absence (or suppression) of nonmagnetic scattering processes connecting states with opposite (or nearly opposite) spin polarization. Moreover, impurity scattering processes coupling zero-energy surface states to bulk nodal quasiparticles are suppressed,
due to the vanishing bulk density of states at zero energy.

\textit{(ii)}
Second, we note that the topological charge (e.g., Eq.~(\ref{eq:GenWno})) can be defined also for gapless topological states with dilute
impurities, for example, by periodically repeating a finite-size disordered system. Such an approach shows that the topological number 
of a disordered nodal noncentrosymmetric superconductor remains nonzero for a large set of quasi-momenta. In passing, we mention 
that this method has recently been
applied to study zero-energy edge states in graphene in the presence of edge disorder~\cite{Wakabayashi2009, Dahal2010}.

\textit{(iii)}  
Third, we discuss the role played by symmetries which restrict the form of the impurity potential. In particular,  chiral symmetry (sublattice symmetry) usually prohibits a large number of scattering channels.
For example, for two-dimensional nodal topological superconductors in symmetry class BDI~\cite{Wong2012}  (and also for graphene~\cite{Akhmerov2008,Wimmer2010})  it was shown that on a given edge all localized zero-energy states have the same chirality number, i.e., all zero-energy edge states are simultaneous eigenstates of the chiral symmetry operator $\mathcal{S}$ with the same eigenvalue $+1$ (or $-1$). Since on-site impurities do not break chiral symmetry, the total chirality number of a given edge remains unchanged in the presence of impurities, and hence the total number of  zero-energy edge states is not altered by disorder.
For noncentrosymmetric superconductors (Sec.~\ref{sec:NCS}), on the other hand, chiral symmetry is less restrictive, since on a given surface there are zero-energy states with both chirality numbers, $+1$ and $-1$.

In summary, the above arguments suggest that the zero-energy surface states of gapless topological phases are at least partially robust against disorder. A more detailed investigation of the influence of  disorder on the topological surface states is left  for future work.

\section{Summary and discussion}
\label{Discussion}

In this paper we have developed a general and unified classification of topologically stable Fermi surfaces in (semi-)metals
and nodal lines in superconductors in terms of discrete symmetries and spatial dimension.
Using K-theory arguments, we have shown that stable Fermi surfaces can be classified in a similar manner
as fully gapped topological states  (see Table~\ref{single fermi}).
The remarkable topological properties of these stable Fermi surfaces and nodal lines manifest themselves in the appearance of
protected zero-energy states at the boundary of the system.
In fact, the presence
of  topological boundary modes is directly linked to the topological structure of the bulk wavefunctions  via a bulk-boundary correspondence. 
Depending on the case, these topological surface states form either one- or two-dimensional flat bands, or linearly dispersing Dirac or Majorana states (see Sec.~\ref{bulkBndCorr}). It should be possible to detect these different surface states using various experimental probes, such as angle-resolved photoemission measurements,  scanning tunneling spectroscopy, or
angle-resolved thermal transport measurements \cite{Matsuda2006}.
 
To illustrate the general principles of the classification scheme, we have examined a few concrete examples, 
specifically of stable nodal lines in three-dimensional noncentrosymmetric superconductors. These nodal superconductors
exhibit dispersionless zero-energy surface states (i.e., surface flat bands) of topological origin.
An important experimental fingerprint of these zero-energy flat bands is a zero-bias peak in the surface density of states, which depends 
strongly on the surface orientation. This dependence can be used as a probe of  the  pairing symmetry
and the bulk nodal structure  of the superconductor. We have also studied
the stability of the surface flat bands against disorder and time-reversal symmetry breaking perturbations.

For some of the gapless topological states listed in Table~\ref{single fermi} (and Table~\ref{total-table}), physical realizations are known and their surface states have been studied extensively. E.g., zero-energy boundary modes have been experimentally observed  in  graphene  
\cite{Niimi2005,Kobayashi2005,Niimi2006} and in $d_{x^2-y^2}$-wave high-temperature superconductors \cite{Kashiwaya1995, Alff1997, Wei1998}. For other entries in Table~\ref{single fermi}, candidate materials have been proposed, but the topological surface state have not yet been conclusively observed. This includes nodal noncentrosymmetric superconductors (class DIII or AIII, $d_{k}=1$),  with the candidate materials CePt$_3$Si~\cite{PhysRevLett.92.027003}, Li$_2$Pt$_3$B~\cite{yuan06}, and BiPd~\cite{Mintu2012}, and 
Weyl semi-metals (class A, $d_{k}=2$), which might be realized in Y$_2$Ir$_2$O$_7$ \cite{WanTurnerAschvin,PhysRevB.85.045124}
or in HgCr$_2$Se$_4$~\cite{PhysRevLett.107.186806}.
Finally, there are also other entries in Table~\ref{single fermi} for which no physical realization is as yet known.
We hope that the results of this paper will spur further experimental investigates of these
interesting gapless topological states.

\paragraph{Note added.}

Upon completion of the manuscript,
we became aware of Ref.\ \cite{Zhao12},
where a similar classification of nodal systems is discussed. 
The result of Ref.\ \cite{Zhao12} agrees with the classification given in Table~\ref{single fermi}.

\ack

APS is grateful to the Kavli Institute for Theoretical Physics for hospitality during the preparation of this work.
The authors thank M.\ Sigrist for discussions.
SR thanks the organizers of ICTP Workshop School (July 2011)
``Workshop and School on Topological Aspects of Condensed Matter Physics'',
where he presented the results of this manuscript.

\section*{References}

\def\urlprefix{}
 \def\url#1{}
\bibliographystyle{iopart-num}
\bibliography{fermi_surf}

\providecommand{\newblock}{}
\begin{thebibliography}{100}
\expandafter\ifx\csname url\endcsname\relax
  \def\url#1{{\tt #1}}\fi
\expandafter\ifx\csname urlprefix\endcsname\relax\def\urlprefix{URL }\fi
\providecommand{\eprint}[2][]{\url{#2}}

\bibitem{konig07}
{K{\"o}nig} M, {Wiedmann} S, {Br{\"u}ne} C, {Roth} A, {Buhmann} H, {Molenkamp}
  L~W, {Qi} X~L and {Zhang} S~C 2007 {\em Science\/} {\bf 318} 766

\bibitem{hsiehNature08}
{Hsieh} D, {Qian} D, {Wray} L, {Xia} Y, {Hor} Y~S, {Cava} R~J and {Hasan} M~Z
  2008 {\em Nature\/} {\bf 452} 970

\bibitem{hasanKaneReview10}
{Hasan} M~Z and {Kane} C~L 2010 {\em Rev. Mod. Phys.\/} {\bf 82} 3045

\bibitem{qiZhangReview10}
Qi X~L and Zhang S~C 2011 {\em Rev. Mod. Phys.\/} {\bf 83} 1057
  \urlprefix\url{http://link.aps.org/doi/10.1103/RevModPhys.83.1057}

\bibitem{HasanMooreReview11}
{Hasan} M~Z and {Moore} J~E 2011 {\em Annu. Rev. Condens. Matter Phys.\/} {\bf
  2} 55

\bibitem{ryuNJP10}
{Ryu} S, {Schnyder} A~P, {Furusaki} A and {Ludwig} A~W~W 2010 {\em New J.
  Phys.\/} {\bf 12} 065010

\bibitem{zhangPRB08}
Qi X~L, Hughes T~L and Zhang S~C 2008 {\em Phys. Rev. B\/} {\bf 78} 195424
  \urlprefix\url{http://link.aps.org/doi/10.1103/PhysRevB.78.195424}

\bibitem{fuKanePRL08}
Fu L and Kane C~L 2008 {\em Phys. Rev. Lett.\/} {\bf 100} 096407
  \urlprefix\url{http://link.aps.org/doi/10.1103/PhysRevLett.100.096407}

\bibitem{heikkila2011a}
{Heikkil{\"a}} T~T and {Volovik} G~E 2011 {\em JETP Letters\/} {\bf 93} 59

\bibitem{heikkila2011b}
{Heikkil{\"a}} T~T, {Kopnin} N~B and {Volovik} G~E 2011 {\em JETP Letters\/}
  {\bf 94} 233

\bibitem{volovik2011c}
{Volovik} G~E 2011 {\em arXiv:1111.4627\/}

\bibitem{nakada96}
Nakada K, Fujita M, Dresselhaus G and Dresselhaus M~S 1996 {\em Phys. Rev. B\/}
  {\bf 54} 17954 
  \urlprefix\url{http://link.aps.org/doi/10.1103/PhysRevB.54.17954}

\bibitem{fujita96}
{Fujita} M, {Wakabayashi} K, {Nakada} K and {Kusakabe} K 1996 {\em J. Phys.
  Soc. Jpn.\/} {\bf 65} 1920

\bibitem{neto2009}
Castro~Neto A~H, Guinea F, Peres N~M~R, Novoselov K~S and Geim A~K 2009 {\em
  Rev. Mod. Phys.\/} {\bf 81} 109
  \urlprefix\url{http://link.aps.org/doi/10.1103/RevModPhys.81.109}

\bibitem{Niimi2005}
Niimi Y, Matsui T, Kambara H, Tagami K, Tsukada M and Fukuyama H 2005 {\em
  Appl. Surf. Sci.\/} {\bf 241} 43
  \urlprefix\url{http://www.sciencedirect.com/science/article/pii/S01694332040%
13601}

\bibitem{Kobayashi2005}
Kobayashi Y, Fukui K~i, Enoki T, Kusakabe K and Kaburagi Y 2005 {\em Phys. Rev.
  B\/} {\bf 71} 193406
  \urlprefix\url{http://link.aps.org/doi/10.1103/PhysRevB.71.193406}

\bibitem{Niimi2006}
Niimi Y, Matsui T, Kambara H, Tagami K, Tsukada M and Fukuyama H 2006 {\em
  Phys. Rev. B\/} {\bf 73} 085421
  \urlprefix\url{http://link.aps.org/doi/10.1103/PhysRevB.73.085421}

\bibitem{huPRL94}
Hu C~R 1994 {\em Phys. Rev. Lett.\/} {\bf 72} 1526
  \urlprefix\url{http://link.aps.org/doi/10.1103/PhysRevLett.72.1526}

\bibitem{ryu2002}
Ryu S and Hatsugai Y 2002 {\em Phys. Rev. Lett.\/} {\bf 89} 077002
  \urlprefix\url{http://link.aps.org/doi/10.1103/PhysRevLett.89.077002}

\bibitem{kashiwaya2000}
{Kashiwaya} S and {Tanaka} Y 2000 {\em Rep. Prog. Phys.\/} {\bf 63} 1641

\bibitem{Kashiwaya1995}
Kashiwaya S, Tanaka Y, Koyanagi M, Takashima H and Kajimura K 1995 {\em Phys.
  Rev. B\/} {\bf 51} 1350
  \urlprefix\url{http://link.aps.org/doi/10.1103/PhysRevB.51.1350}

\bibitem{Alff1997}
Alff L, Takashima H, Kashiwaya S, Terada N, Ihara H, Tanaka Y, Koyanagi M and
  Kajimura K 1997 {\em Phys. Rev. B\/} {\bf 55} R14757
  \urlprefix\url{http://link.aps.org/doi/10.1103/PhysRevB.55.R14757}

\bibitem{Wei1998}
Wei J~Y~T, Yeh N~C, Garrigus D~F and Strasik M 1998 {\em Phys. Rev. Lett.\/}
  {\bf 81} 2542--2545
  \urlprefix\url{http://link.aps.org/doi/10.1103/PhysRevLett.81.2542}

\bibitem{volovikJETP2011}
{Volovik} G~E 2011 {\em JETP Letters\/} {\bf 93} 66

\bibitem{tsutsumi2011}
Tsutsumi Y, Ichioka M and Machida K 2011 {\em Phys. Rev. B\/} {\bf 83} 094510
  \urlprefix\url{http://link.aps.org/doi/10.1103/PhysRevB.83.094510}

\bibitem{WangFa2012}
Wang F and Lee D~H 2012 {\em Phys. Rev. B\/} {\bf 86} 094512
  \urlprefix\url{http://link.aps.org/doi/10.1103/PhysRevB.86.094512}

\bibitem{sato06}
Sato M 2006 {\em Phys. Rev. B\/} {\bf 73} 214502
  \urlprefix\url{http://link.aps.org/doi/10.1103/PhysRevB.73.214502}

\bibitem{beriPRB2010}
B\'eri B 2010 {\em Phys. Rev. B\/} {\bf 81} 134515
  \urlprefix\url{http://link.aps.org/doi/10.1103/PhysRevB.81.134515}

\bibitem{Schnyder2010}
Schnyder A~P and Ryu S 2011 {\em Phys. Rev. B\/} {\bf 84} 060504
  \urlprefix\url{http://link.aps.org/doi/10.1103/PhysRevB.84.060504}

\bibitem{brydon2010}
Brydon P~M~R, Schnyder A~P and Timm C 2011 {\em Phys. Rev. B\/} {\bf 84} 020501
  \urlprefix\url{http://link.aps.org/doi/10.1103/PhysRevB.84.020501}

\bibitem{Schnyder2011}
Schnyder A~P, Brydon P~M~R and Timm C 2012 {\em Phys. Rev. B\/} {\bf 85} 024522
  \urlprefix\url{http://link.aps.org/doi/10.1103/PhysRevB.85.024522}

\bibitem{satoPRB2011}
Sato M, Tanaka Y, Yada K and Yokoyama T 2011 {\em Phys. Rev. B\/} {\bf 83}
  224511 \urlprefix\url{http://link.aps.org/doi/10.1103/PhysRevB.83.224511}

\bibitem{yadaPRB2011}
Yada K, Sato M, Tanaka Y and Yokoyama T 2011 {\em Phys. Rev. B\/} {\bf 83}
  064505 \urlprefix\url{http://link.aps.org/doi/10.1103/PhysRevB.83.064505}

\bibitem{tanakaPRL2010}
Tanaka Y, Mizuno Y, Yokoyama T, Yada K and Sato M 2010 {\em Phys. Rev. Lett.\/}
  {\bf 105} 097002
  \urlprefix\url{http://link.aps.org/doi/10.1103/PhysRevLett.105.097002}

\bibitem{satoPRL2010}
Sato M and Fujimoto S 2010 {\em Phys. Rev. Lett.\/} {\bf 105} 217001
  \urlprefix\url{http://link.aps.org/doi/10.1103/PhysRevLett.105.217001}

\bibitem{Horava}
Ho\ifmmode~\check{r}\else \v{r}\fi{}ava P 2005 {\em Phys. Rev. Lett.\/} {\bf
  95} 016405
  \urlprefix\url{http://link.aps.org/doi/10.1103/PhysRevLett.95.016405}

\bibitem{volovik2007}
Volovik G~E 2007 {\em Lect. Notes Phys.\/} {\bf 718} 31
  \urlprefix\url{http://dx.doi.org/10.1007/3-540-70859-6-3}

\bibitem{BernardKimLeClair1}
Bernard D, Kim E~A and LeClair A 2012 {\em Phys. Rev. B\/} {\bf 86} 205116
  \urlprefix\url{http://link.aps.org/doi/10.1103/PhysRevB.86.205116}

\bibitem{BernardKimLeClair2}
{Buceta} R~C and {Hansmann} D 2012 {\em J. Phys. A: Math. Theor.\/} {\bf 45}
  435202

\bibitem{schnyderPRB08}
Schnyder A~P, Ryu S, Furusaki A and Ludwig A~W~W 2008 {\em Phys. Rev. B\/} {\bf
  78} 195125 \urlprefix\url{http://link.aps.org/doi/10.1103/PhysRevB.78.195125}

\bibitem{zirnbauerMathPhys96}
Zirnbauer M~R 1996 {\em J. Math. Phys.\/} {\bf 37} 4986
  \urlprefix\url{http://link.aip.org/link/?JMP/37/4986/1}

\bibitem{altlandZirnbauer97}
Altland A and Zirnbauer M~R 1997 {\em Phys. Rev. B\/} {\bf 55} 1142
  \urlprefix\url{http://link.aps.org/doi/10.1103/PhysRevB.55.1142}

\bibitem{heinznerCommMath05}
Heinzner P, Huckleberry A and Zirnbauer M~R 2005 {\em Commun. Math. Phys.\/}
  {\bf 257} 725

\bibitem{kitaev09}
Kitaev A 2009 {\em AIP Conf. Proc.\/} {\bf 1134} 22
  \urlprefix\url{http://link.aip.org/link/?APC/1134/22/1}

\bibitem{Teo-Kane2010}
Teo J~C~Y and Kane C~L 2010 {\em Phys. Rev. B\/} {\bf 82} 115120
  \urlprefix\url{http://link.aps.org/doi/10.1103/PhysRevB.82.115120}

\bibitem{WanTurnerAschvin}
Wan X, Turner A~M, Vishwanath A and Savrasov S~Y 2011 {\em Phys. Rev. B\/} {\bf
  83} 205101 \urlprefix\url{http://link.aps.org/doi/10.1103/PhysRevB.83.205101}

\bibitem{Witten:1998cd}
{Witten} E 1998 {\em JHEP\/} {\bf 9812} 019

\bibitem{DbranesTI}
Ryu S and Takayanagi T 2010 {\em Phys. Lett. B\/} {\bf 693} 175 
  \urlprefix\url{http://www.sciencedirect.com/science/article/pii/S03702693100%
09524}

\bibitem{DbranesTIPRD}
Ryu S and Takayanagi T 2010 {\em Phys. Rev. D\/} {\bf 82} 086014
  \urlprefix\url{http://link.aps.org/doi/10.1103/PhysRevD.82.086014}

\bibitem{Freedman2010}
Freedman M, Hastings M~B, Nayak C, Qi X~L, Walker K and Wang Z 2011 {\em Phys.
  Rev. B\/} {\bf 83} 115132
  \urlprefix\url{http://link.aps.org/doi/10.1103/PhysRevB.83.115132}

\bibitem{2009AIPC.1134...10S}
{Schnyder} A~P, {Ryu} S, {Furusaki} A and {Ludwig} A~W~W 2009 {\em AIP Conf.
  Proc.\/} {\bf 1134} 10--21

\bibitem{Grinevich88}
Grinevich P and Volovik G 1988 {\em J. Low Temp. Phys.\/} {\bf 72} 371
  \urlprefix\url{http://dx.doi.org/10.1007/BF00682148}

\bibitem{zhangPRX12}
Wang Z and Zhang S~C 2012 {\em Phys. Rev. X\/} {\bf 2} 031008
  \urlprefix\url{http://link.aps.org/doi/10.1103/PhysRevX.2.031008}

\bibitem{burkovPRL2011}
Burkov A~A and Balents L 2011 {\em Phys. Rev. Lett.\/} {\bf 107} 127205
  \urlprefix\url{http://link.aps.org/doi/10.1103/PhysRevLett.107.127205}

\bibitem{burkovPRB2011}
Burkov A~A, Hook M~D and Balents L 2011 {\em Phys. Rev. B\/} {\bf 84} 235126
  \urlprefix\url{http://link.aps.org/doi/10.1103/PhysRevB.84.235126}

\bibitem{PhysRevB.85.045124}
Witczak-Krempa W and Kim Y~B 2012 {\em Phys. Rev. B\/} {\bf 85} 045124
  \urlprefix\url{http://link.aps.org/doi/10.1103/PhysRevB.85.045124}

\bibitem{PhysRevLett.107.186806}
Xu G, Weng H, Wang Z, Dai X and Fang Z 2011 {\em Phys. Rev. Lett.\/} {\bf 107}
  186806 \urlprefix\url{http://link.aps.org/doi/10.1103/PhysRevLett.107.186806}

\bibitem{PhysRevB.86.115208}
Singh B, Sharma A, Lin H, Hasan M~Z, Prasad R and Bansil A 2012 {\em Phys. Rev.
  B\/} {\bf 86} 115208
  \urlprefix\url{http://link.aps.org/doi/10.1103/PhysRevB.86.115208}

\bibitem{murakamiNJP07}
Murakami S 2007 {\em New J. Phys.\/} {\bf 9} 356
  \urlprefix\url{http://stacks.iop.org/1367-2630/9/i=9/a=356}

\bibitem{murakami08}
Murakami S and Kuga S~i 2008 {\em Phys. Rev. B\/} {\bf 78} 165313
  \urlprefix\url{http://link.aps.org/doi/10.1103/PhysRevB.78.165313}

\bibitem{schnyderPRB10}
Schnyder A~P, Brydon P~M~R, Manske D and Timm C 2010 {\em Phys. Rev. B\/} {\bf
  82} 184508 \urlprefix\url{http://link.aps.org/doi/10.1103/PhysRevB.82.184508}

\bibitem{Nielsen198120}
Nielsen H and Ninomiya M 1981 {\em Nuclear Physics B\/} {\bf 185} 20
  \urlprefix\url{http://www.sciencedirect.com/science/article/pii/055032138190%
3618}

\bibitem{Gurarie}
Essin A~M and Gurarie V 2011 {\em Phys. Rev. B\/} {\bf 84} 125132
  \urlprefix\url{http://link.aps.org/doi/10.1103/PhysRevB.84.125132}

\bibitem{paananenDahm12}
{Paananen} T and {Dahm} T 2012 {\em arXiv:1210.4422\/}

\bibitem{volovikBOOKS}
Volovik G~E 2003 {\em {The Universe in a Helium Droplet}\/} (Clarendon Press ;
  Oxford University Press)
  \urlprefix\url{http://www.amazon.com/exec/obidos/redirect?tag=citeulike07-20%
\&path=ASIN/0198507828}

\bibitem{roy2008}
{Roy} R 2008 {\em arXiv: 0803.2868\/}

\bibitem{schnyderPRL09}
Schnyder A~P, Ryu S and Ludwig A~W~W 2009 {\em Phys. Rev. Lett.\/} {\bf 102}
  196804 \urlprefix\url{http://link.aps.org/doi/10.1103/PhysRevLett.102.196804}

\bibitem{kaneMelea05a}
Kane C~L and Mele E~J 2005 {\em Phys. Rev. Lett.\/} {\bf 95} 226801
  \urlprefix\url{http://link.aps.org/doi/10.1103/PhysRevLett.95.226801}

\bibitem{kaneMelea05b}
Kane C~L and Mele E~J 2005 {\em Phys. Rev. Lett.\/} {\bf 95} 146802
  \urlprefix\url{http://link.aps.org/doi/10.1103/PhysRevLett.95.146802}

\bibitem{fuKane07}
Fu L and Kane C~L 2007 {\em Phys. Rev. B\/} {\bf 76} 045302
  \urlprefix\url{http://link.aps.org/doi/10.1103/PhysRevB.76.045302}

\bibitem{moorePRB07}
Moore J~E and Balents L 2007 {\em Phys. Rev. B\/} {\bf 75} 121306
  \urlprefix\url{http://link.aps.org/doi/10.1103/PhysRevB.75.121306}

\bibitem{royPRB09}
Roy R 2009 {\em Phys. Rev. B\/} {\bf 79} 195321
  \urlprefix\url{http://link.aps.org/doi/10.1103/PhysRevB.79.195321}

\bibitem{vollhardt1990superfluid}
Vollhardt D and Woelfle P 1990 {\em The Superfluid Phases Of Helium 3\/}
  (Taylor \& Francis)
  \urlprefix\url{http://books.google.de/books?id=t0Rw75gMuwIC}

\bibitem{Nishikubo2011}
{Nishikubo} Y, {Kudo} K and {Nohara} M 2011 {\em J. Phys. Soc. Jpn.\/} {\bf 80}
  055002

\bibitem{Goryo2012}
Goryo J, Fischer M~H and Sigrist M 2012 {\em Phys. Rev. B\/} {\bf 86} 100507
  \urlprefix\url{http://link.aps.org/doi/10.1103/PhysRevB.86.100507}

\bibitem{2012arXiv1212.2441B}
{Biswas} P~K, {Luetkens} H, {Neupert} T, {Stuerzer} T, {Baines} C, {Pascua} G,
  {Schnyder} A~P, {Fischer} M~H, {Goryo} J, {Lees} M~R, {Maeter} H, {Brueckner}
  F, {Klauss} H~H, {Nicklas} M, {Baker} P~J, {Hillier} A~D, {Sigrist} M,
  {Amato} A and {Johrendt} D 2012 {\em arXiv: 1212.2441\/}

\bibitem{bauerSigristBook}
Bauer E and Sigrist M 2012 {\em Non-Centrosymmetric Superconductors:
  Introduction and Overview\/} ({\em Lecture Notes in Physics\/} vol 847)
  (Springer Berlin)

\bibitem{Huang08}
{Huang} Y, {Yan} J, {Wang} Y, {Shan} L, {Luo} Q, {Wang} W and {Wen} H~H 2008
  {\em Supercond. Sci. Technol.\/} {\bf 21} 075011

\bibitem{Klimczuk07}
Klimczuk T, Ronning F, Sidorov V, Cava R~J and Thompson J~D 2007 {\em Phys.
  Rev. Lett.\/} {\bf 99} 257004
  \urlprefix\url{http://link.aps.org/doi/10.1103/PhysRevLett.99.257004}

\bibitem{Bonalde2011}
Bonalde I, Kim H, Prozorov R, Rojas C, Rogl P and Bauer E 2011 {\em Phys. Rev.
  B\/} {\bf 84} 134506
  \urlprefix\url{http://link.aps.org/doi/10.1103/PhysRevB.84.134506}

\bibitem{Lue2011}
Lue C~S, Su T~H, Liu H~F and Young B~L 2011 {\em Phys. Rev. B\/} {\bf 84}
  052509 \urlprefix\url{http://link.aps.org/doi/10.1103/PhysRevB.84.052509}

\bibitem{Akazawa04}
{Akazawa} T, {Hidaka} H, {Kotegawa} H, {Kobayashi} T~C, {Fujiwara} T,
  {Yamamoto} E, {Haga} Y, {Settai} R and {{\= O}nuki} Y 2004 {\em J. Phys. Soc.
  Jpn.\/} {\bf 73} 3129

\bibitem{yuan06}
Yuan H~Q, Agterberg D~F, Hayashi N, Badica P, Vandervelde D, Togano K, Sigrist
  M and Salamon M~B 2006 {\em Phys. Rev. Lett.\/} {\bf 97} 017006
  \urlprefix\url{http://link.aps.org/doi/10.1103/PhysRevLett.97.017006}

\bibitem{PhysRevLett.98.047002}
Nishiyama M, Inada Y and Zheng G~q 2007 {\em Phys. Rev. Lett.\/} {\bf 98}
  047002 \urlprefix\url{http://link.aps.org/doi/10.1103/PhysRevLett.98.047002}

\bibitem{Bhanu2011}
Joshi B, Thamizhavel A and Ramakrishnan S 2011 {\em Phys. Rev. B\/} {\bf 84}
  064518 \urlprefix\url{http://link.aps.org/doi/10.1103/PhysRevB.84.064518}

\bibitem{Mintu2012}
Mondal M, Joshi B, Kumar S, Kamlapure A, Ganguli S~C, Thamizhavel A, Mandal
  S~S, Ramakrishnan S and Raychaudhuri P 2012 {\em Phys. Rev. B\/} {\bf 86}
  094520 \urlprefix\url{http://link.aps.org/doi/10.1103/PhysRevB.86.094520}

\bibitem{PhysRevLett.92.027003}
Bauer E, Hilscher G, Michor H, Paul C, Scheidt E~W, Gribanov A, Seropegin Y,
  No\"el H, Sigrist M and Rogl P 2004 {\em Phys. Rev. Lett.\/} {\bf 92} 027003
  \urlprefix\url{http://link.aps.org/doi/10.1103/PhysRevLett.92.027003}

\bibitem{Onuki09}
{Onuki} R, {Sumiyama} A, {Oda} Y, {Yasuda} T, {Settai} R and {{\= O}nuki} Y
  2009 {\em J. Phys.: Condens. Matter\/} {\bf 21} 075703

\bibitem{Sugitani06}
{Sugitani} I, {Okuda} Y, {Shishido} H, {Yamada} T, {Thamizhavel} A, {Yamamoto}
  E, {Matsuda} T~D, {Haga} Y, {Takeuchi} T, {Settai} R and {{\= O}nuki} Y 2006
  {\em J. Phys. Soc. Jpn.\/} {\bf 75} 043703

\bibitem{Kimura05}
Kimura N, Ito K, Saitoh K, Umeda Y, Aoki H and Terashima T 2005 {\em Phys. Rev.
  Lett.\/} {\bf 95} 247004
  \urlprefix\url{http://link.aps.org/doi/10.1103/PhysRevLett.95.247004}

\bibitem{PhysRevB.86.174520}
Dahlhaus J~P, Gibertini M and Beenakker C~W~J 2012 {\em Phys. Rev. B\/} {\bf
  86} 174520 \urlprefix\url{http://link.aps.org/doi/10.1103/PhysRevB.86.174520}

\bibitem{2013arXiv1302.3461S}
{Schnyder} A~P, {Timm} C and {Brydon} P~M~R 2013 {\em arXiv: 1302.3461\/}

\bibitem{2013arXiv1302.3714B}
{Brydon} P~M~R, {Timm} C and {Schnyder} A~P 2013 {\em New J. Phys.\/} {\bf 15}
  045019

\bibitem{2013arXiv1302.1943T}
{Tafti} F~F, {Fujii} T, {Juneau-Fecteau} A, {de Cotret} S~R, {Doiron-Leyraud}
  N, {Asamitsu} A and {Taillefer} L 2013 {\em arXiv: 1302.1943\/}

\bibitem{samokhinAnnals09}
Samokhin K 2009 {\em Ann. Phys.\/} {\bf 324} 2385
  \urlprefix\url{http://www.sciencedirect.com/science/article/pii/S00034916090%
01559}

\bibitem{Wakabayashi2009}
Wakabayashi K, Takane Y, Yamamoto M and Sigrist M 2009 {\em Carbon\/} {\bf 47}
  124
  \urlprefix\url{http://www.sciencedirect.com/science/article/pii/S00086223080%
04971}

\bibitem{Dahal2010}
Dahal H~P, Hu Z~X, Sinitsyn N~A, Yang K and Balatsky A~V 2010 {\em Phys. Rev.
  B\/} {\bf 81} 155406
  \urlprefix\url{http://link.aps.org/doi/10.1103/PhysRevB.81.155406}

\bibitem{Wong2012}
Wong C~L~M, Liu J, Law K~T and Lee P~A 2012 {\em arXiv: 1206.5601\/}

\bibitem{Akhmerov2008}
Akhmerov A~R and Beenakker C~W~J 2008 {\em Phys. Rev. B\/} {\bf 77} 085423
  \urlprefix\url{http://link.aps.org/doi/10.1103/PhysRevB.77.085423}

\bibitem{Wimmer2010}
Wimmer M, Akhmerov A~R and Guinea F 2010 {\em Phys. Rev. B\/} {\bf 82} 045409
  \urlprefix\url{http://link.aps.org/doi/10.1103/PhysRevB.82.045409}

\bibitem{Matsuda2006}
{Matsuda} Y, {Izawa} K and {Vekhter} I 2006 {\em J. Phys.: Condens. Matter\/}
  {\bf 18} R705

\bibitem{Zhao12}
{Zhao} Y~X and {Wang} Z~D 2012 {\em arXiv: 1211.7241\/}

\end{thebibliography}

\end{document}